\documentclass{aa}  

\usepackage{txfonts}
\usepackage[english]{babel}
\usepackage{amsmath}
\usepackage{amssymb}
\newcommand{\ignore}[1]{}
\usepackage{natbib}
\bibpunct{(}{)}{;}{a}{}{,} 
\usepackage{amssymb}
\usepackage{longtable}
\usepackage{graphicx}

\newcommand{\mearth}{M_\oplus}

\def\ms{\hbox{\,m\,s$^{-1}$}}         
\def\m2s2{\hbox{\,m$^{2}$\,s$^{-2}$}} 

\usepackage{color}

\begin{document}

   \title{HADES RV program with HARPS-N at the TNG\thanks{Based on: observations made with the Italian {\it Telescopio Nazionale Galileo} (TNG), operated on the island of La Palma by the INAF - {\it Fundaci{\'o}n Galileo Galilei} at the {\it Roche de Los Muchachos} Observatory of the {\it Instituto de Astrof{\'i}sica de Canarias} (IAC); photometric observations made with the APACHE array located at the Astronomical Observatory of the Aosta Valley; photometric observations made with the robotic telescope
APT2 (within the EXORAP program) located at Serra La Nave on Mt. Etna.} \\IX. A super-Earth around the M dwarf \object{Gl\,686} \thanks{RV data (Table \ref{TabA.1}) and APACHE and EXORAP photometry are only available in electronic form at the CDS via anonymous ftp to cdsarc.u-strasbg.fr (130.79.128.5) or via http://cdsweb.u-strasbg.fr/cgi-bin/qcat?J/A+A/
} }

\titlerunning{The HADES RV program}

   \author{L. Affer\inst{1}
          \and
          M. Damasso\inst{2}
          \and
          G. Micela\inst{1}
          \and
          E. Poretti\inst{3,4}
          \and
          G. Scandariato\inst{5}
          \and
          J. Maldonado\inst{1}
          \and
          A. F. Lanza\inst{5}
          \and
          E. Covino\inst{6}
          \and
          A. Garrido Rubio\inst{1}
          \and
          J. I. Gonz{\'a}lez Hern{\'a}ndez\inst{7,8}
          \and
          R. Gratton\inst{9}
          \and
          G. Leto\inst{5}
          \and
          A. Maggio\inst{1}
          \and
          M. Perger\inst{10,11}   
          \and
          A. Sozzetti\inst{2}     
          \and
          A. Su{\'a}rez Mascare{\~n}o\inst{7,12}
          \and
          A. S. Bonomo\inst{2}
          \and
          F. Borsa\inst{3}
          \and
          R. Claudi\inst{9}
          \and
          R. Cosentino\inst{4}
	  \and
          S. Desidera\inst{9}
	  \and
	  P. Giacobbe\inst{2}
          \and
          E. Molinari\inst{13}
          \and
          M. Pedani\inst{4}
          \and
          M. Pinamonti\inst{2}
          \and
          R. Rebolo\inst{7,8}
          \and
          I. Ribas\inst{10,11}
          \and
          B. Toledo-Padr{\'o}n\inst{7,8}          
          }
 
   \institute{INAF - Osservatorio Astronomico di Palermo, Piazza del Parlamento 1, 90134 Palermo, Italy\\
              \email{laura.affer@inaf.it}
              \and
              INAF - Osservatorio Astrofisico di Torino, via Osservatorio 20, 10025 Pino Torinese, Italy
              \and
              INAF - Osservatorio Astronomico di Brera, via E. Bianchi 46, 23807 Merate (LC), Italy
              \and
              INAF - Fundaci{\'o}n Galileo Galilei, Rambla Jos{\'e} Ana Fernandez P{\'e}rez 7, 38712, Bre{\~n}a Baja, TF, Spain
              \and
              INAF - Osservatorio Astrofisico di Catania, via S. Sofia 78, 95123 Catania, Italy
              \and
              INAF - Osservatorio Astronomico di Napoli, Salita Moiariello, 16, 80131 Napoli, Italy
              \and
              Instituto de Astrof{\'{i}}sica de Canarias, 38205 La Laguna, Tenerife, Spain 
              \and
              Universidad de La Laguna, Dpto Astrof{\'{i}}sica, 38206 La Laguna, Tenerife, Spain
              \and
              INAF - Osservatorio Astronomico di Padova, Vicolo dell'Osservatorio 5, 35122, Padova, Italy
              \and
              Institut de Ci{\'e}ncies de l'Espai (ICE,CSIC), Campus UAB, Carrer de Can Magrans s/n, 08193 Bellaterra, Spain
              \and
              Institut d'Estudis Espacials de Catalunya (IEEC), 08034 Barcelona, Spain
              \and
              Observatoire Astronomique de l'Universit{\'e} de Gen{\`e}ve, 1290, Versoix, Switzerland
              \and
              INAF - Osservatorio Astronomico di Cagliari, Viale della Scienza, 5, 09047 Selargius, Cagliari, Italy           
              }


 
  \abstract
   {}
   {The HArps-n red Dwarf Exoplanet Survey\thanks{http://www.oact.inaf.it/exoit/EXO-IT/Projects/Entries/2011/12/27\_GAPS.html}      (HADES) is providing a major contribution to the widening of the current statistics of low-mass planets, through the in-depth analysis of precise radial-velocity (RV) measurements in a narrow range of spectral sub-types. Using the HARPS-N spectrograph we reach the precision needed to detect small planets with a few earth masses. Our survey is mainly focused on the M-dwarf population of the northern hemisphere. }
   {As part of that program, we obtained RV measurements of \object{Gl\,686}, an M1 dwarf at d = 8.2 pc. These measurements show a dispersion much in excess of their internal errors. The analysis of data obtained within an intensive observing campaign demonstrates that the excess dispersion is due to a coherent signal with a period of 15.53 d. Almost simultaneous photometric observations were carried out within the APACHE and EXORAP programs to characterize the stellar activity and to distinguish periodic variations related to activity from signals due to the presence of planetary companions, complemented also with ASAS photometric data. We used a Bayesian framework to estimate the orbital parameters and the planet minimum mass, and to properly treat the activity noise. We took advantage of the available RV measurements for this target from other observing campaigns. The analysis of the RV composite time series from the HIRES, HARPS, and HARPS-N spectrographs, consisting of 198 measurements taken over 20 yr, enabled us to address the nature of periodic signals and also to characterize stellar physical parameters (mass, temperature, and rotation).}
   {We report the discovery of a super-Earth orbiting at a distance of 0.092 AU from the host star \object{Gl\,686}. The planet has a minimum mass of $7.1\pm\, 0.9$\, $\mearth$ and an orbital period of $15.532\pm\, 0.002\, d$. The analysis of the activity indexes, of the correlated noise through a Gaussian process framework, and of the photometry provides an estimate of the stellar rotation period at 37 d, and highlights the variability of the spot configuration during the long timespan covering 20 yr. The observed periodicities around 2000 d likely point to the existence of an activity cycle.}
   {}

  \keywords{techniques: radial velocities - techniques: photometric - methods: data analysis - stars: individual: Gl686 - instrumentation: spectrographs - planets and satellites: detection
              }

   \maketitle
%

\section{INTRODUCTION}
Very low-mass stars are very promising targets for planet search programs, and are particularly useful for the discovery of super-Earths/Earths located in their
habitable zone (HZ). Their detection is, in principle, possible with the existing spectrographs of highest radial-velocity (RV) precision, but it is challenging due to stellar activity which introduces correlated and uncorrelated noise into RV curves (RV jitter). On the other hand, the lower masses of M dwarfs result in a higher Doppler wobble RV amplitude for a
given planetary mass, and the low luminosities of M dwarfs imply that the boundaries of their HZ are located at short separations \citep[typically between 0.02 AU and 0.2 AU, see, e.g.,][]{man07}, making habitable rocky planets more easy to detect with present-day observing facilities than those around more massive stars \citep[e.g.,][]{tuo14,cha07}.
The most recent evidence gathered by RV surveys and ground-based as well as space-borne transit search programs points toward the ubiquitousness of low-mass companions with small radii around M dwarfs \citep[e.g.,][]{soz13,cro15,dre15}. 
The number of confirmed planets is steadily growing and in recent years many planetary systems hosting Neptune-mass planets \citep[e.g.,][]{how14,ast15}, super-Earths \citep[e.g.,][]{del13,rib18}, and Earth-mass planets \citep[e.g.,][]{ang16,gil17,ast17a}  have been
reported. Nevertheless, the frequency statistics of low-mass planets hosted by low-mass stars remains poorly constrained. Several studies have attempted to quantify the abundance of rocky planets in close orbits and in the habitable zones of M dwarfs, but the uncertainties are still large, making it important to continue adding new planets to the sample \citep{bon13,gai13,dre15}.\\

We present high-precision, high-resolution spectroscopic measurements of the bright M1 dwarf \object{Gl\,686}, gathered with the HARPS-N spectrograph (\citealp{cos12}) on the Telescopio Nazionale Galileo (TNG) as part of the HArps-n red Dwarf Exoplanet Survey (HADES), described in detail in \citet{aff16}. The HADES collaboration has already produced many valuable results, regarding the statistics, activity, and characterization of M stars \citep{per17a,mal17,sca17,mal15,sua18}, and has led to the discovery of several planets \citep{aff16,sua17,per17b,pin18}.
\\ For the study of \object{Gl\,686} we also took advantage of archive observations  with the HARPS and HIRES spectrographs, as part of other observing campaigns. In Sect.~\ref{sec:sec2} we summarize the atmospheric, physical, and kinematic properties of \object{Gl\,686}. In Sect.~\ref{sec:sec3} we describe the datasets and the RV analysis and discuss the effect of stellar activity for the time series of each instrument and for the composite time series. Section~\ref{sec:sec4} presents the analysis and results of a multi-site photometric monitoring campaign. In Sect.~\ref{sec:sec5} we describe the Bayesian analysis and the model selection and derive the system parameters. In Sect.~\ref{sec:sec6} we discuss our results, and we summarize our findings and conclude in Sect.~\ref{sec:sec7}.

%
\begin{table}[h]
\centering
\caption{Stellar parameters for the star \object{Gl\,686} from the analysis of the HARPS-N spectra using the technique in Maldonado et al. (2015) (upper part); the log\,$R'_{\mathrm{HK}}$ value was calculated by \citealp{sua18}. The lower part of the table lists coordinates, V and K magnitudes, parallax, proper motions and space velocities are indicated. }             
\label{table:1}      
\begin{tabular}{ c c }     
\hline\hline      
Parameter$^{(5)}$ & \object{Gl\,686}\\\hline
Spectral Type  & M1 \\  
$T_{\rm eff}$ [K]   & 3663$\pm$68   \\
$[{\rm Fe}/{\rm H}]$ [dex]    & -0.30$\pm$0.09  \\
Mass [M$_{\odot}$] & 0.42$\pm$0.05   \\
Radius [R$_{\odot}$] & 0.42$\pm$0.05   \\
log\,$g$ [cgs] & 4.83$\pm$0.04   \\
Luminosity [L$_*$/L$_{\odot}$] & 0.028$\pm$0.006   \\
v sin\,$i$ [km s$^{-1}$] & 1.01$\pm$0.80   \\
log\,$R'_{\mathrm{HK}}$$^{(1)}$ &  -5.42$\pm$0.05 \\
\hline
$\alpha$ (J2000) & $17^{h}$:37$^{m}$:$52.8^{s}$ \\
$\delta$ (J2000) & +$18^{o}$:$35'$:$21''$ \\
$V_{mag}$$^{(2)}$& 9.577 \\
$K_{mag}$$^{(3)}$& 5.572 \\
$\pi$[mas]$^{(4)}$& 122.56$\pm$0.03\\
$\mu_{\alpha}$[mas/yr]$^{(4)}$& 926.74$\pm$0.05\\
$\mu_{\delta}$[mas/yr]$^{(4)}$& 984.70$\pm$0.07\\
$U_{LSR}$ [km s$^{-1}$]$^{(5)}$ & -24.4$\pm$0.4 \\
$V_{LSR}$ [km s$^{-1}$]$^{(5)}$ &  40.2$\pm$0.6 \\   
$W_{LSR}$ [km s$^{-1}$]$^{(5)}$ & -13.9$\pm$0.4 \\    
S [km s$^{-1}$]$^{(5)}$ &  49.0$\pm$0.6\\\hline\hline                  
\end{tabular}
\begin{flushleft}
References. $^{(1)}$ \citet{sua18}; $^{(2)}$ \citet{koe10}; $^{(3)}$ \citet{cut03}; $^{(4)}$ \citet{gaia16,gaia18}; $^{(5)}$ This work (see text).
\end{flushleft}
\end{table}
%


\section{Stellar properties of \object{Gl\,686}}\label{sec:sec2}
\object{Gl\,686} is a variable dwarf at a distance of 8 pc from the Sun. The rotation period of \object{Gl\,686} was estimated by \citet{sua18} within the HADES program using 9 HARPS-N and 20 HARPS spectra (see \citealt{sua18} for details on the method). Though a direct determination was not possible, the expected rotation period is 70\,$\pm$\,12 d. \citet{ast17b} estimated a rotation period of 55 d for \object{Gl\,686} using only the HARPS data.\\ 
Accurate stellar parameters were determined using the empirical relations by \citet{mal15}\footnote{http://www.astropa.inaf.it/\textasciitilde{}jmaldonado/Msdlines.html} on the same HARPS-N spectra 
used in the present work to derive RVs (see Table~\ref{table:1}). \\
Stars presently near the Sun may come from a wide range of Galactic locations. Therefore, stellar space velocity represents an important clue to the origin of a star in the Galaxy. The accurate $\it Gaia$ parallax \citep{gaia16,gaia18} combined with the proper motions and the stellar RV makes it possible to derive reliable space velocities for \object{\object{Gl\,686}}. All kinematic
data and stellar properties are listed in Table~\ref{table:1}. We calculated the probabilities
that the star belongs to a specific population: thick disk (TD), thin disk (D), or stellar halo (H), following the method used by \citet{ben04}. 
On account of these probabilities, we find for
\object{Gl\,686} a thick-disk to thin-disk probability ratio of $TD/D=0.07$, implying that the star is clearly identified as a 
thin-disk object (typical threshold for assignment to thin disk being $TD/D\,$<$\,  0.1$, \citealp{ben04}).

\begin{figure}
\centering
\includegraphics[width=9cm]{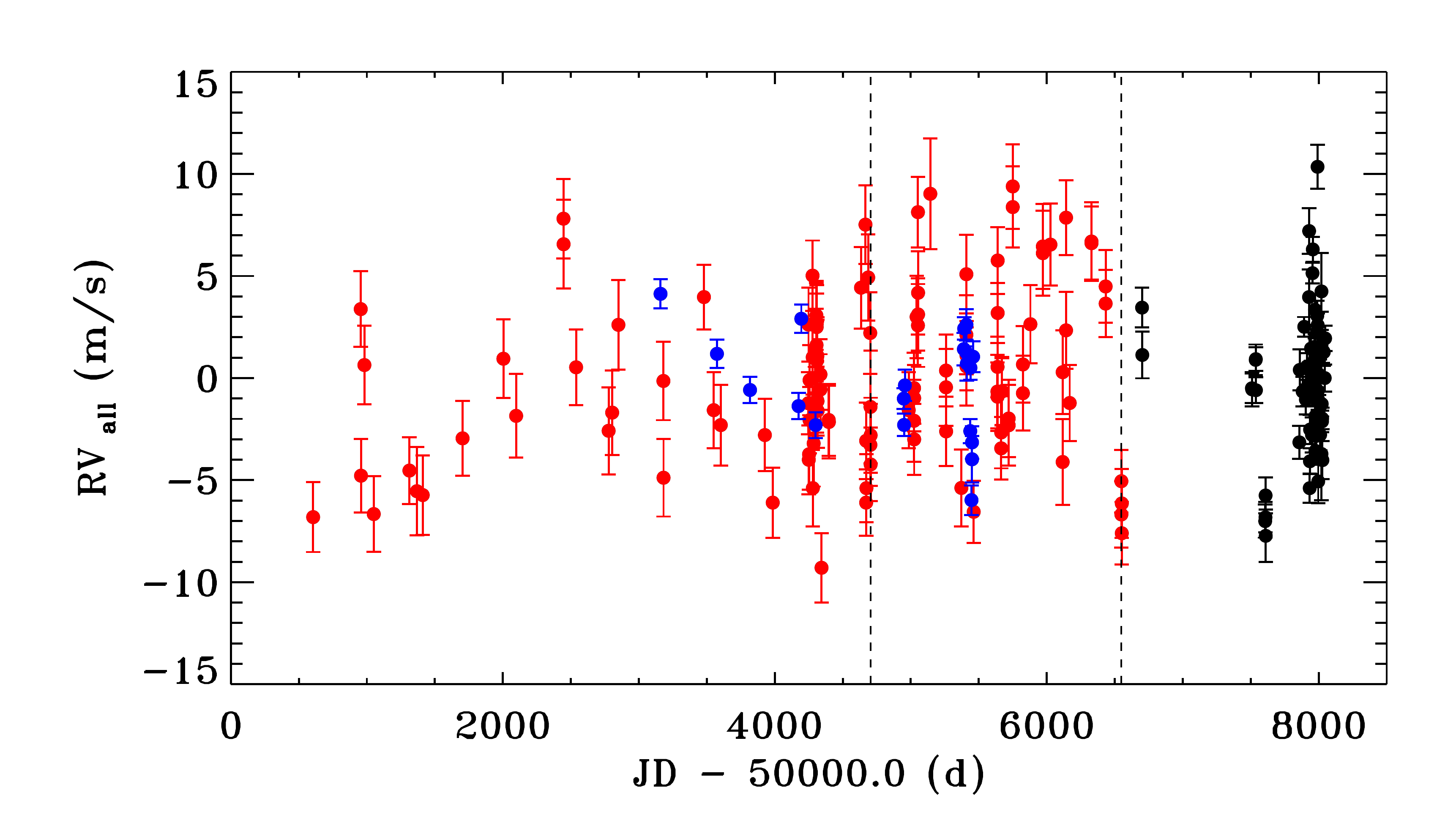}
\caption{Composite RV time series from HIRES (red), HARPS (blue), and HARPS-N (black) for \object{Gl\,686}, after offset correction. The two vertical lines indicate the separation between the different $\it seasons$,  each composed of 66 data points.}
\label{fig:1}
\end{figure}

\section{Datasets and analysis}\label{sec:sec3}
\subsection{Data}

Over the last 20 yrs, \object{Gl\,686} has been monitored by three different surveys and instruments:
from BJD = 2\,450\,604.9 (June 5, 1997) to BJD = 2\,456\,551.8 (September 16, 2013), using the optical echelle HIRES spectrograph on the Keck-I telescope within the survey carried out by the Lick-Carnegie Exoplanet Survey Team (LCES), obtaining 114 spectra at high resolution (archive public data). For HIRES data we used the RVs measured by \citet{but17}, taking into account the recent correction by \citet{tal18}; from BJD = 2\,453\,159.7 (June 3, 2004) to BJD = 2\,455\,458.5 (September 19, 2010), using the optical echelle HARPS spectrograph on the ESO La Silla telescope (\citealp{may03}), obtaining 20 spectra at high resolution (archive public data); from BJD = 2\,456\,700.7 (February 12, 2014) to BJD = 2\,458\,047.3 (October 20, 2017), using the HARPS-N
spectrograph on the TNG telescope, as part of the HADES program.\\ The spectra were obtained at high resolution with exposure times of 15 minutes and average signal-to-noise ratio (S/N) of 75 at 5500 \AA. Of the 64  HARPS-N epochs, 49 were
obtained within the GAPS observing program and 15 within the Spanish observing program. Observations were gathered without the simultaneous
Th-Ar calibration, which is usually used to correct for instrumental drifts during the night. \citet{per17a} demonstrated that we can account for the lack of inter-night instrumental drift adding quadratically a mean instrumental drift of 1 m s$^{-1}$ to the RV uncertainties. This choice avoided the contamination of the Ca II H and K lines, which are particularly important for the stellar activity analysis of M dwarfs \citep{gia89,for09,lov11}. 

The information on datasets, instruments, and surveys used in the present work is summarized in Table~\ref{table:2}. The RV time series for HARPS and HARPS-N are provided in Table~\ref{TabA.1}.

\begin{table}[h]
\scriptsize
\centering
\caption{Information on the datasets and instruments used in the present work.}             
\label{table:2}            
\begin{tabular}{ l l r l r }     
\hline\hline      
Survey& Telescope/ & Resolving & Time-span  & \#Data\\
& Instrument &  Power&$\Delta$T (d)&points\\\hline
HADES & TNG/HARPS-N & 115\,000 & 1347 & 64\\
LCES  & Keck-I/HIRES & 60\,000 & 5947 & 114\\
HMDS$^{(*)}$  & ESO/HARPS & 115\,000 & 2299 & 20\\\hline                 
\end{tabular}
\begin{flushleft}
$^{(*)}$: The HARPS M dwarf sample; program IDs: 072.C-0488, 183.C-0437.
\end{flushleft}
\end{table}

\begin{figure}
\centering
\includegraphics[width=9cm]{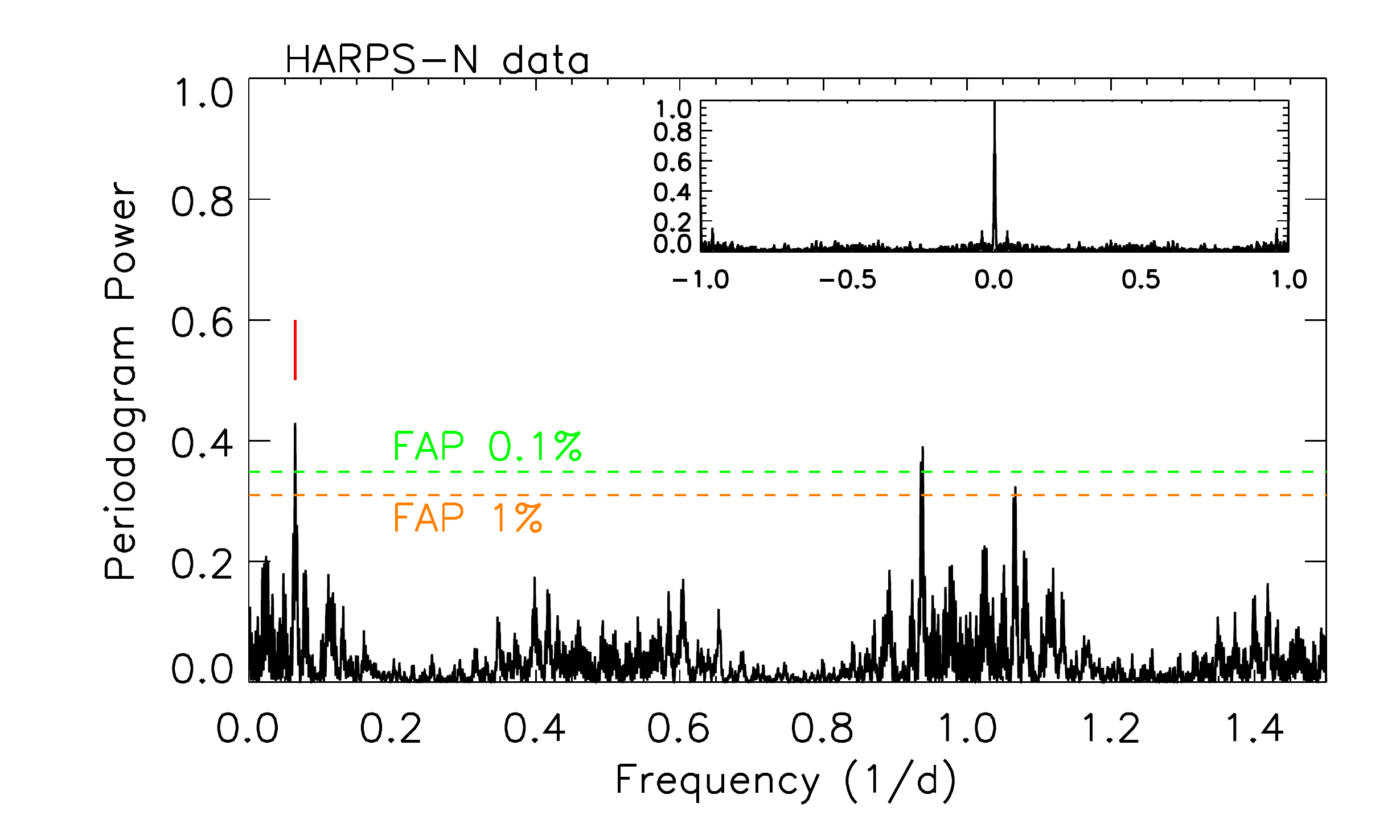}
\includegraphics[width=9cm]{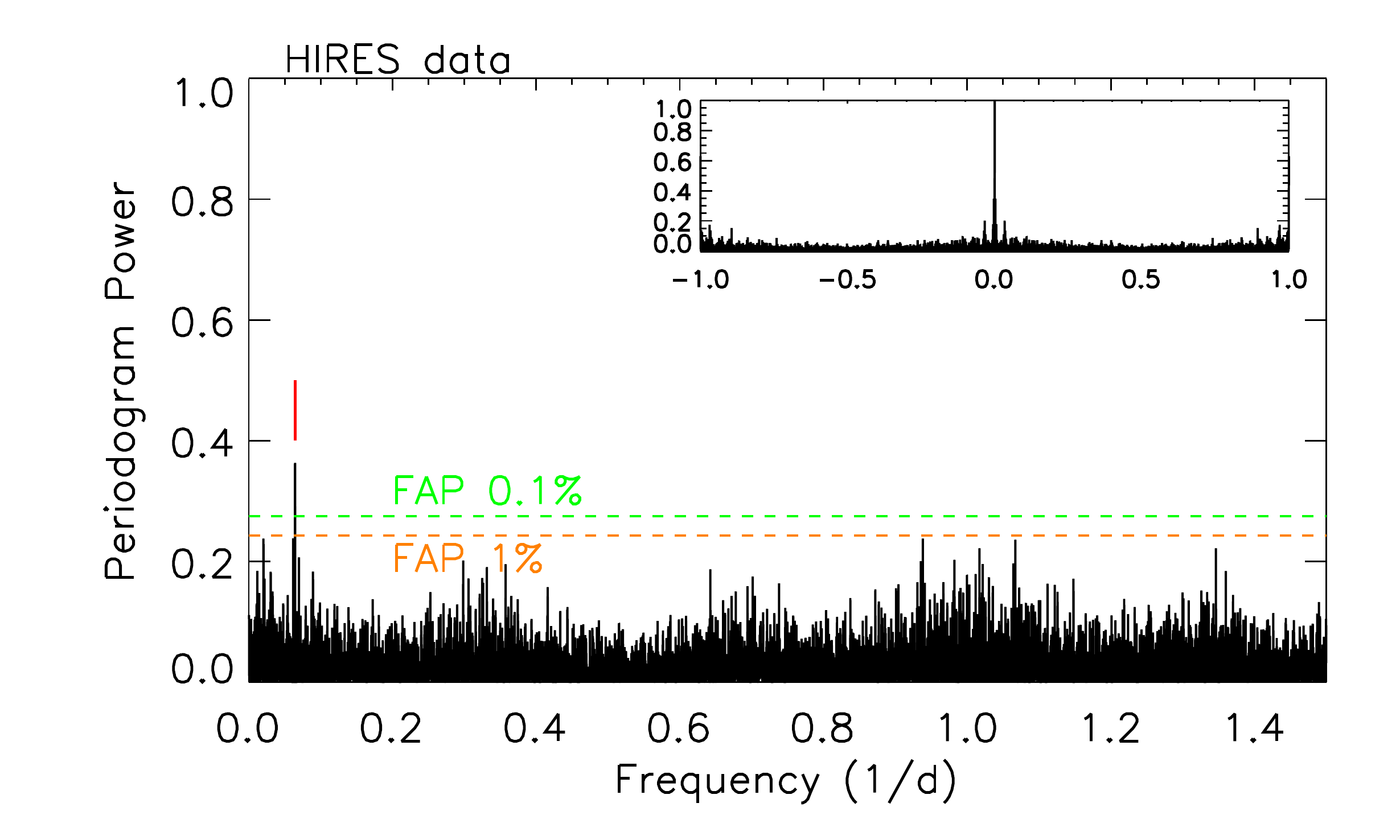}
\includegraphics[width=9cm]{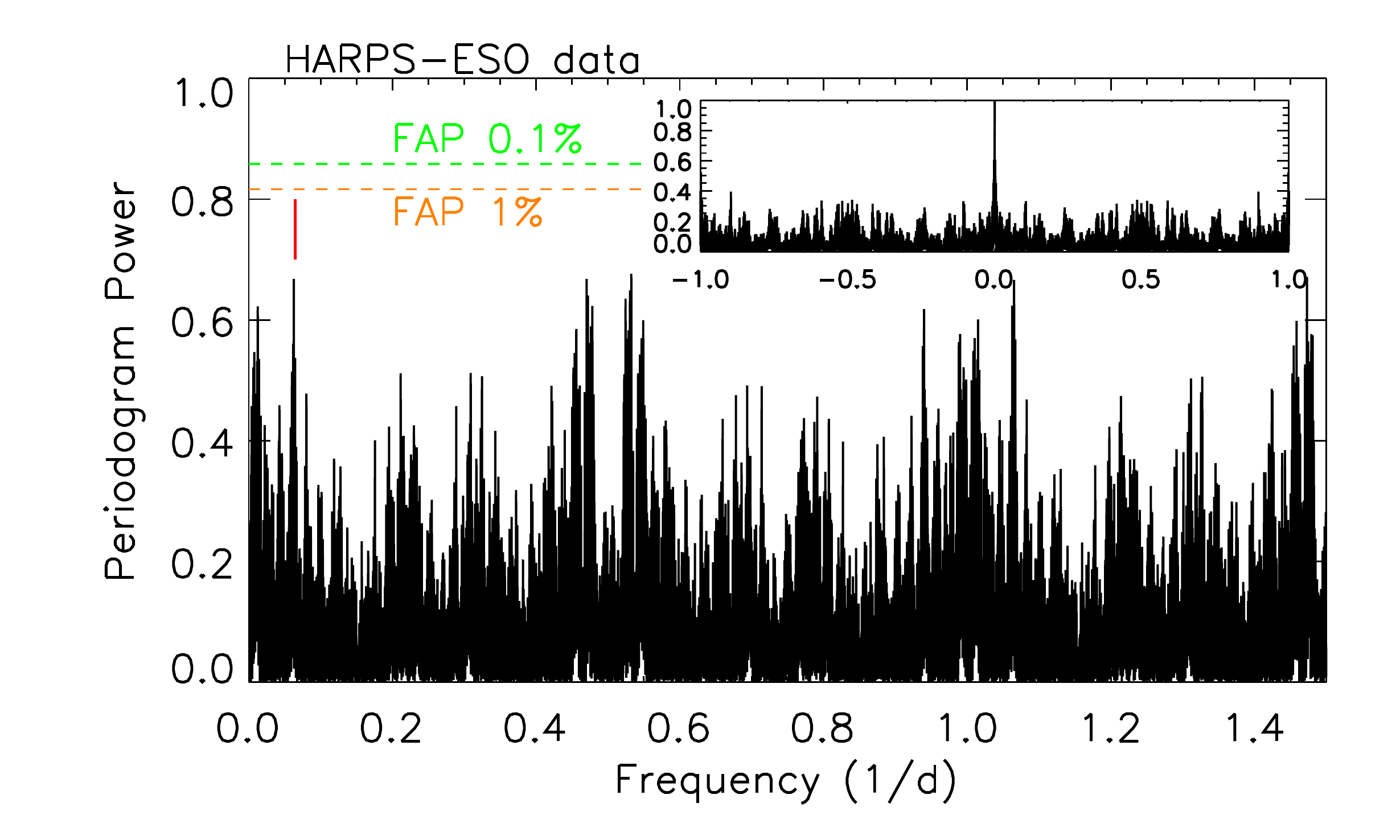}
\caption{Power spectrum of the HARPS-N, HIRES, HARPS data, from top to bottom, with the spectral window in insets. The period at 15.5 d is indicated with a red mark.}
\label{fig:2}
\end{figure}

\subsection{RV period search}\label{3.2}
We performed the same analysis on the three datasets individually, and then again on the composite dataset, including RV measurements from the three spectrographs.
Joining the three different datasets, we obtained a total of 198 data points (114 HIRES - 20 HARPS - 64 HARPS-N) spanning more than 7400
days, with HIRES and HARPS data overlapping. When we consider the RV data of the three different spectrographs HIRES, HARPS, and HARPS-N, we need to compensate for the instrument offset of the different measurements. This is done with the IDL program {\it RVlin.pro} \citep{wri09}. Once the offset is calculated, the RV measurements of two instruments (here HIRES and HARPS) can be corrected for this offset and be brought to the same zero-point as the HARPS-N measurements (Fig.\,\ref{fig:1}). \\
HARPS-N data reduction and spectral extraction were performed using the Data Reduction Software \citep[DRS v3.7,][]{lov07}. Relative velocities were measured by matching the spectra with a high S/N template obtained by co-adding the spectra of the target, as implemented in the TERRA pipeline \citep{ang12}. This strategy provides a better RV accuracy when applied to M-dwarfs, as shown in \citet{per17a}. We employed the same approach to measure the RVs also for HARPS data. \\ The TERRA pipeline performs the correction for the secular acceleration \citep{kur03}, and the RV dataset from the Keck archive was already corrected for the secular acceleration of the host star (0.34 m~s$^{-1}$\,yr$^{-1}$).\\
The first step of the RV data analysis consists in identifying significant periodic signals in the data. Pre-whitening is a commonly used tool for finding multi-periodic signals in time series data. With this method we find sequentially the dominant Fourier components in the time series and remove them. The pre-whitening procedure was applied to the RV data using the Generalized Lomb-Scargle (GLS) periodogram
algorithm \citep{zec09} and the program $\it Period04$ \citep{len04},
for an independent test. We calculated the false alarm probabilities (FAP) of detection using 10\,000 bootstrap randomization \citep{end01} of the original RV time series. 
Once a significant peak was located at a given period, the corresponding sinusoidal function was adjusted and removed. The process was repeated several times until no significant peak remained. The two methods yielded the same results in
terms of extracted frequencies from our RV time series. \\
$\mathbf{HARPS-N\, -}$ The RVs show a root mean square (RMS) dispersion of 3.16 m s$^{-1}$ and a mean formal error of  0.82 m s$^{-1}$ (1.3 m s$^{-1}$, adding quadratically the mean instrumental drift of 1 m s$^{-1}$ to the RV uncertainties) . The GLS periodogram indicates as significant (FAP 1\%) a period of $15.52 \pm\,0.01$\,d, and the residuals time series has a RMS of 2.42 m s$^{-1}$ and a significant period of $36.7 \pm\,0.1$\,d. The time series obtained eliminating these two significant periods has a RMS of 2.0 m s$^{-1}$ and a period of $12.95 \pm\,0.02$\,d, which is nonsignificant at the 1\% FAP threshold.\\
$\mathbf{HARPS\, -}$ The RVs show a RMS dispersion of 2.52 m s$^{-1}$ and a mean error of  1.22 m s$^{-1}$. The GLS periodogram indicates a period of $15.96 \pm\,0.01$\,d, the residuals time series has a RMS of 1.41 m s$^{-1}$ and a period of $4.4947 \pm\,0.0007$\,d; both periods are nonsignificant. \\
$\mathbf{HIRES\, -}$ The RVs show a RMS dispersion of 4.09 m s$^{-1}$ and a mean error of  1.84 m s$^{-1}$. The GLS periodogram indicates a significant period of $15.53 \pm\,0.003$\,d, and the residuals time series has a RMS of 3.3 m s$^{-1}$ and a significant period of $49.09 \pm\,0.04$\,d.  The time series obtained eliminating these two significant periods has a RMS of 2.8 m s$^{-1}$ and a nonsignificant period of $30.3 \pm\,0.02$\,d.\\
The GLS periodogram of the three datasets is shown in Fig.\,\ref{fig:2}, the 0.1\% and 1\% FAP
levels are shown as dashed lines, and the window functions are plotted in insets for each instrument.\\

\begin{figure}
\centering
\includegraphics[width=9cm]{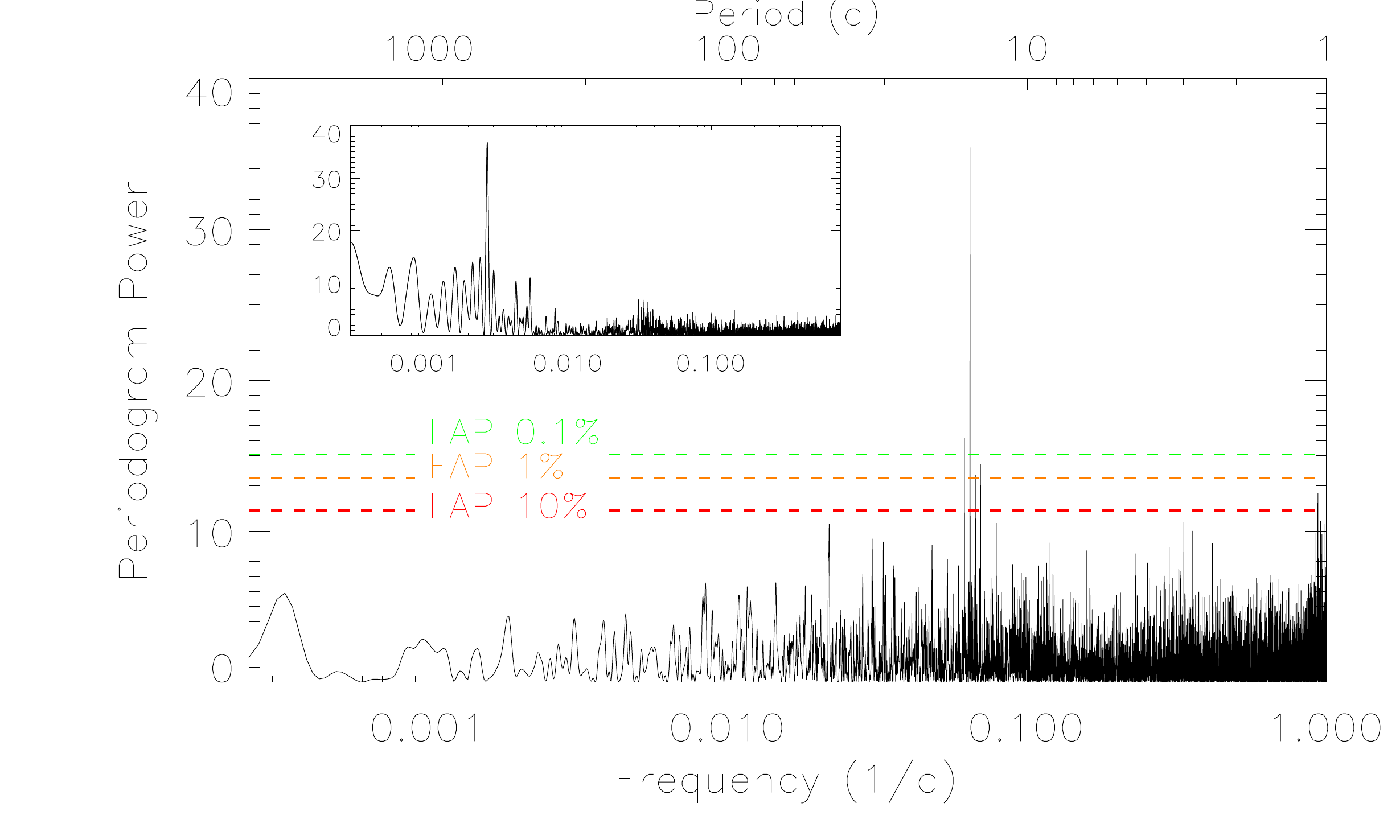}
\caption{GLS periodogram of the complete dataset (HIRES+HARPS+HARPS-N). The dashed lines indicate 0.1\%, 1\%, and 10\% levels of FAP. The spectral window is shown in the inset and the highest peak is related to the one-year alias.}
\label{fig:3}
\end{figure}

\noindent $\mathbf{HIRES+HARPS+HARPS-N\, -}$ The RVs show a RMS dispersion of 3.45 m s$^{-1}$ and a mean error of  1.51 m s$^{-1}$,  the GLS periodogram
(with a period search extended to 3\,700\,d, due to the long timespan of the composite observations) indicates a significant period of
$15.531 \pm\,0.001$\,d with a very high spectral power, and the residuals time series has a RMS of 3.07 m s$^{-1}$ and a significant period of
$45.95 \pm\,0.03$\,d.  The time series obtained eliminating these two significant periods has a RMS of 2.9 m s$^{-1}$ and a significant
period of $31.35 \pm\,0.01$\,d. The GLS periodogram of the complete dataset is shown in Fig.\,\ref{fig:3}; the 0.1\%, 1\%, and 10\% FAP
levels are shown as dashed lines.\\

\begin{figure}
\includegraphics[width=9cm]{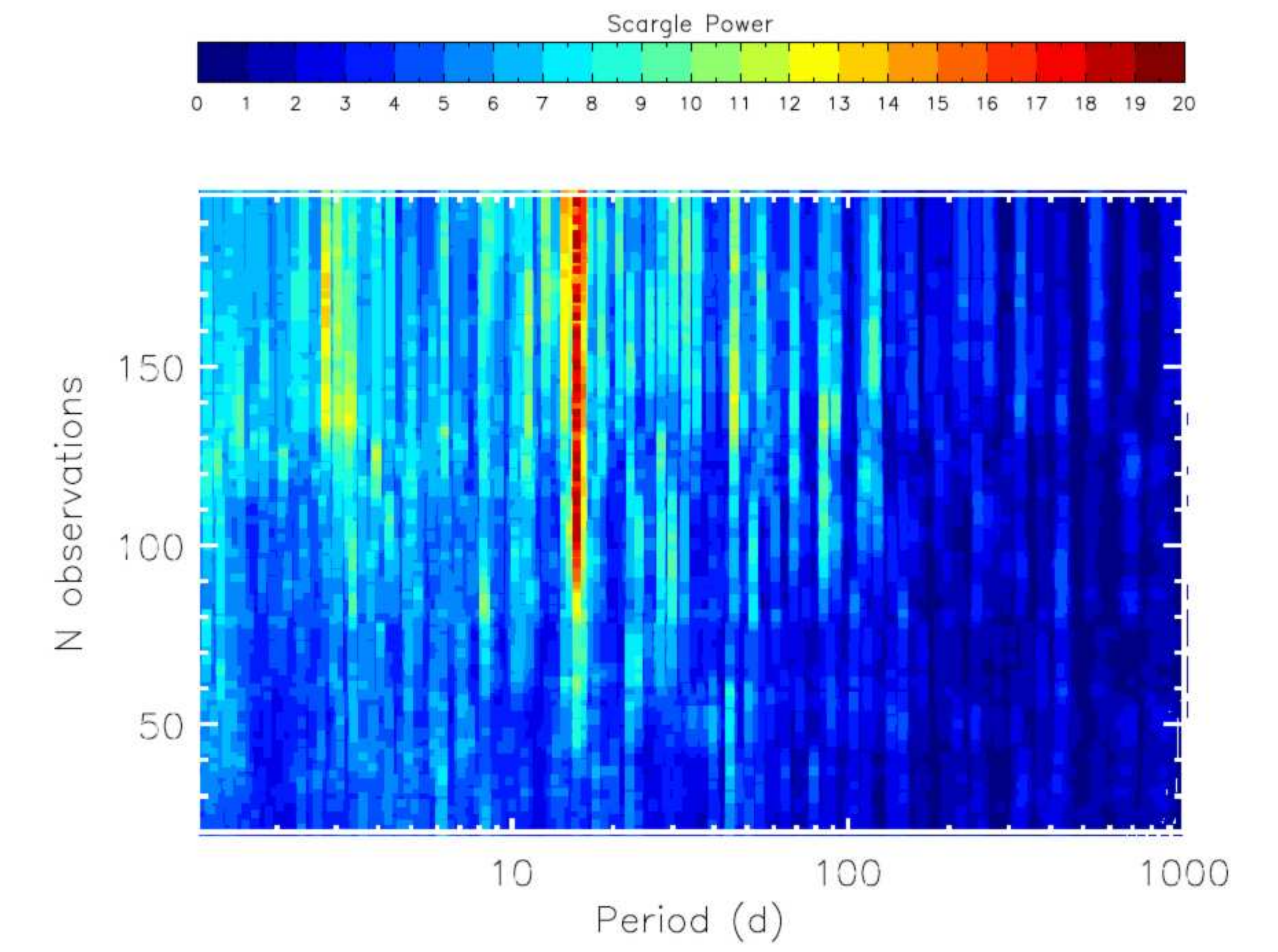}
\centering
\caption{Periodogram power of the inspected periods as a function of the number of observations. The most significant period (the reddest one) at 15.5 d is clearly visible with a high power for N$_{obs}$ $>$\,80, as well as a period around 3 d with a moderate Scargle power, which increases for N$_{obs}$ $>$\,160 and then decreases again as well as a period between 40 and 50 d.}
\label{fig:4}
\end{figure}

To visualize the cumulative contribution of data points to the significance of the detected frequencies, we calculated the GLS periodograms increasing the number of observations \citep{mor17} from 20 to N$_{tot}$, adding one observation at a time (following the exact order of data acquisition); the result is shown in Fig.\,\ref{fig:4}. We compute the power corresponding to each of the periods in the range 1.2 d to P$_{Max}$ = (T$_{BJD_Max}$-T$_{BJD_Min}$)/2.0 for each periodogram obtained. Each horizontal slice of Fig.\,\ref{fig:4} gives power (indicated by the color scale) versus period for the periodogram corresponding to the observation number indicated on the vertical axis. Figure\,\ref{fig:4} is a 2D representation of the technique applied to the complete \object{Gl\,686} dataset (HARPS-N+HARPS+HIRES).\\

\begin{figure}
\centering
\includegraphics[width=9cm]{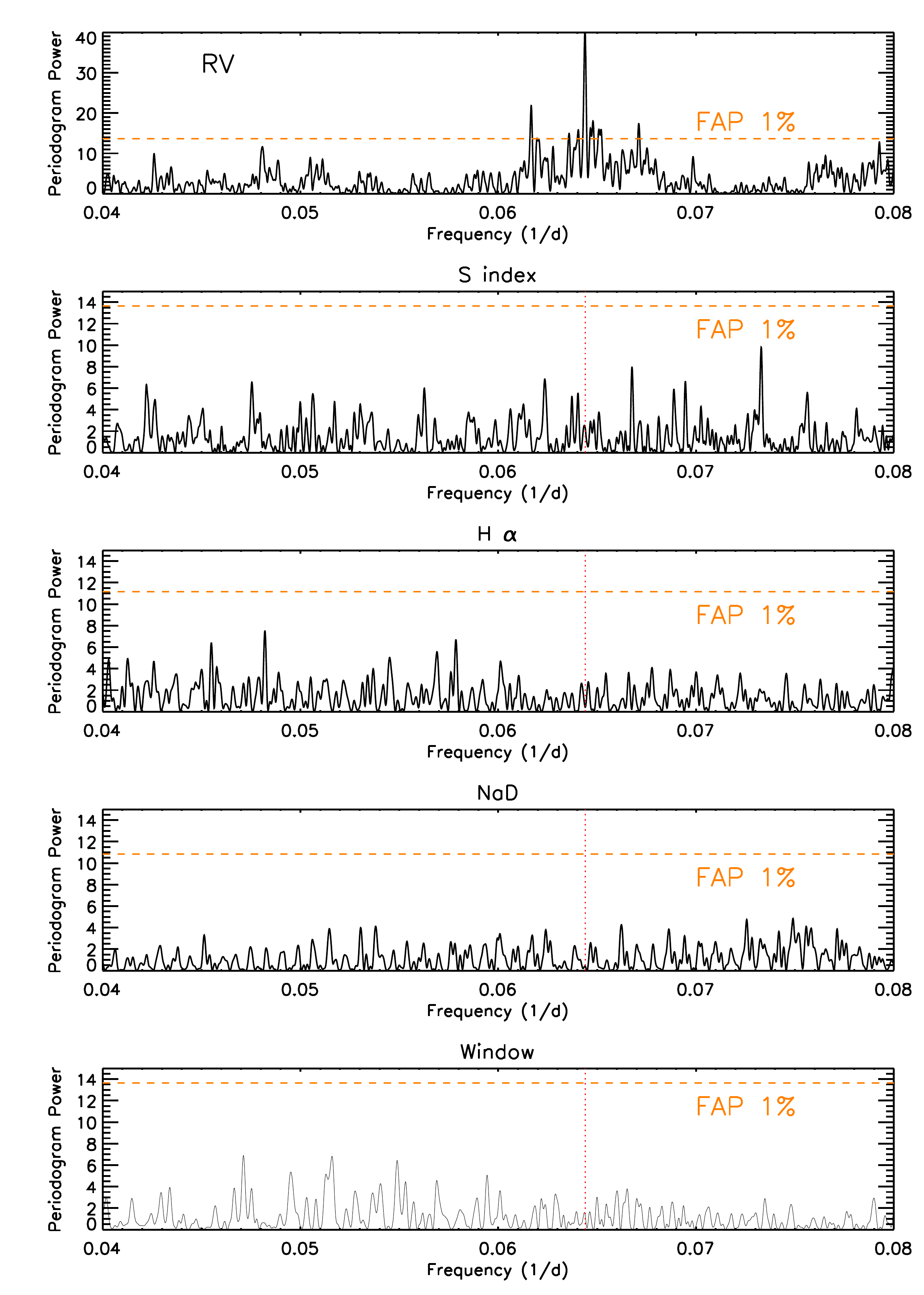}
\caption{GLS periodograms around the period of 15.5 d (from top to bottom): RV (HIRES+HARPS+HARPS-N), Sindex (HIRES+HARPS+HARPS-N), H$\alpha$ (HARPS+HARPS-N), NaD (HARPS+HARPS-N) and window function. The 15.5-d period is indicated in each periodogram, as the red dashed line.}
\label{fig:5}
\end{figure}

We also performed a different analysis using the iterative sine-wave-fitting least-squares method \citep{van71}. This approach is used in the
asteroseismic analysis of multiperiodic pulsating stars to minimize the subtle effects of pre-whitening, such as power exchange between a
signal frequency and an alias of another signal frequency. After each detection, only the frequency values were introduced as known
parameters in the new search. In this way, their amplitudes and phases were recalculated for each new trial frequency, always subtracting
the exact amount of signal for any known periodicity. This new analysis again confirmed the previously determined frequencies. \\

\subsection{Aliases: connection to window function peaks}\label{3.3}
A careful analysis of the spectral window, following the methods of \citet{daw10}, ruled out that the peak at 15.5 d in the periodograms is an artifact due to the combination of the activity with the spectral window (or to the time sampling alone). \\
By taking a closer look at the RV and window function periodograms (Fig.\,\ref{fig:2}), it is clearly visible that the window function does not show significant peaks at the same frequency as the RV periodogram. In particular, since the window function is superimposed at every real frequency, overplotting the window function at the most prominent peaks found in the RV periodogram allows us to see how the ancillary peaks (on both sides) of the real frequency are caused purely by the sampling function (in particular yearly aliases).

\subsection{Stellar activity}\label{3.4}
To investigate RV variability against stellar activity, we make use of spectral indexes based on Ca II H and K ($S$-index), H$\alpha,$ and NaD lines obtained as results of the TERRA pipeline for HARPS-N and HARPS data, and $S$-index and H$\alpha$ available from the Keck archive, for this target (Fig.\,\ref{fig:A1}). The index based on NaD lines was defined following the receipt of \citet{mal15} for Ca II H and K lines. \\
The average Ca II H and K activity index $R'_{\mathrm{HK}}$, measured by \citet{sua18} (Table~\ref{table:1}), indicates a low chromospheric activity level for \object{Gl\,686}.\\
\noindent $\it{HARPS-N\, -}$ The GLS periodogram for H$\alpha$ time series (Fig.\, \ref{fig:A3}), indicates a significant period of $37.05 \pm\,0.07$\,d and no significant peaks for NaD; the significant periodicity observed for the $S$\,index is related to the one-year alias. \\
\noindent $\it{HARPS\, -}$ None of the peaks found in the GLS periodogram of activity indexes of HARPS data are significant. \\
\noindent $\it{HIRES\, -}$ The GLS periodogram indicates a significant period of $45.35 \pm\,0.05$\,d for H$\alpha$ time series
(Fig.\, \ref{fig:A3}) and a significant period of $38.00 \pm\,0.02$\,d for the $S$-index.\\

\begin{figure}
\includegraphics[width=9cm]{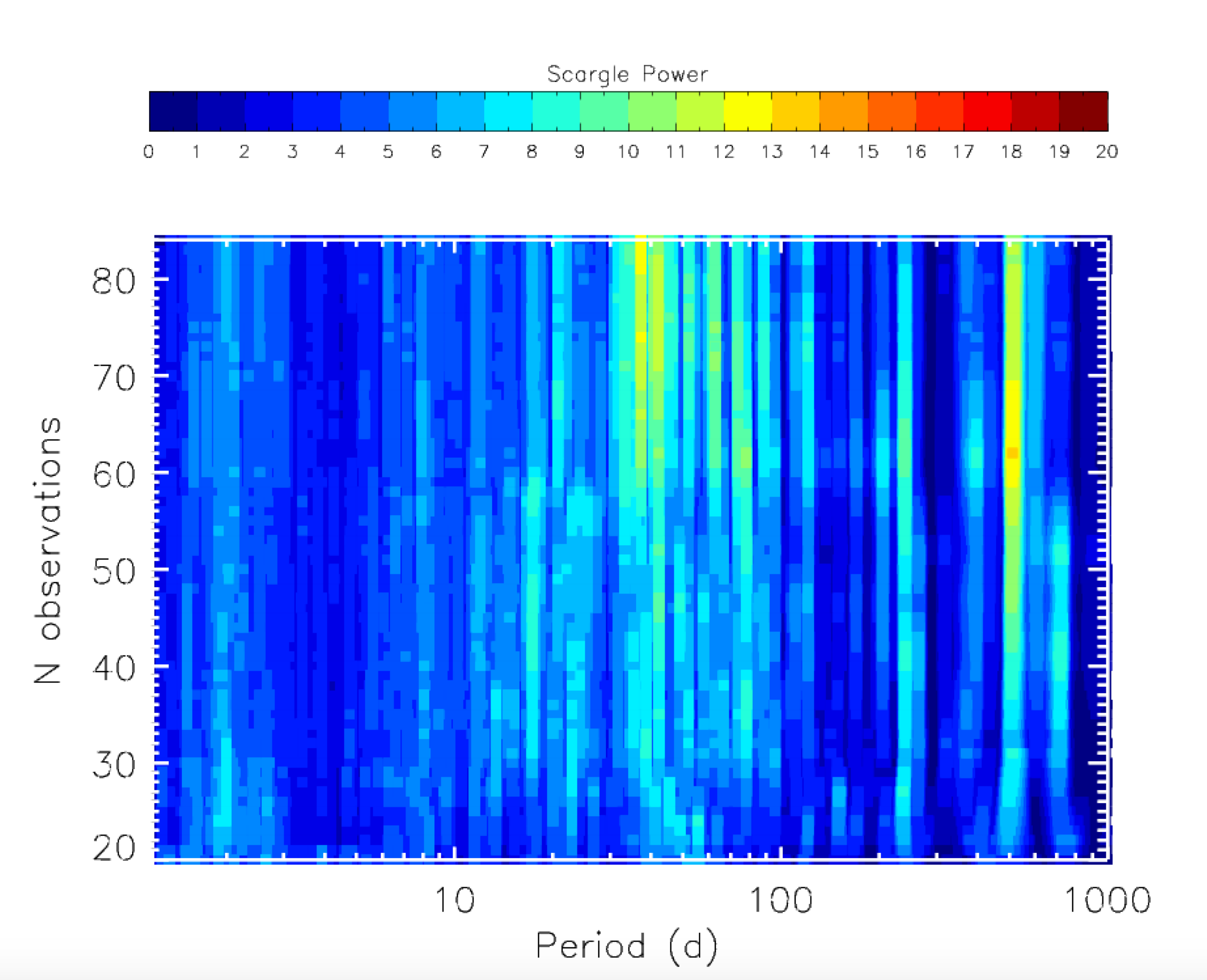}
\includegraphics[width=9cm]{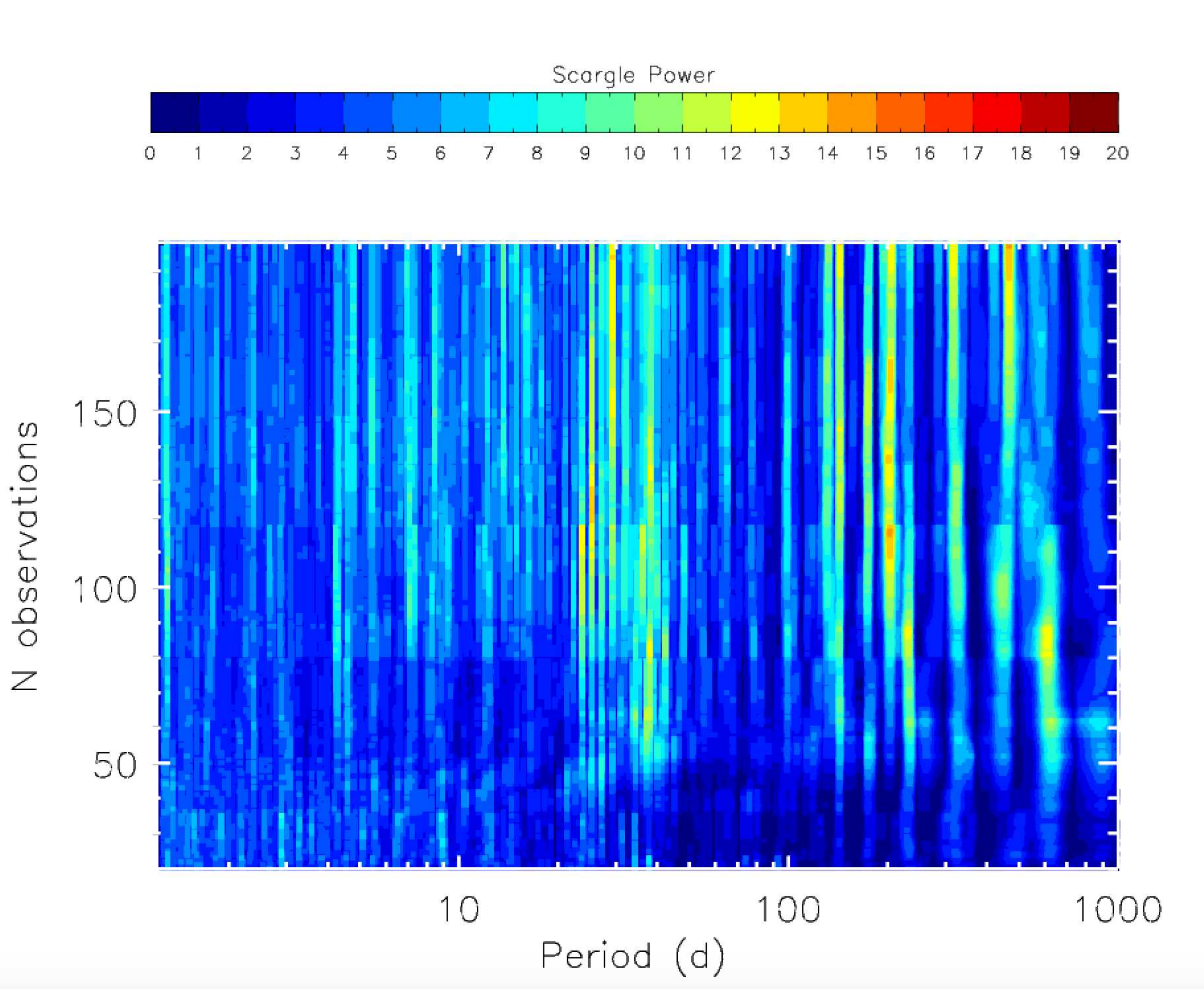}
\centering
\caption{Periodogram power of the inspected periods as a function of the number of observations for the H$\alpha$ (top panel) and $S$-index (bottom panel) time series. There are several significant periods larger than 100 d, probably due to long trends and a series of significant periods of between 27 and 42 d for $S$\,index and around 40 d for H$\alpha$.}
\label{fig:6}
\end{figure}

We performed the activity analysis also for the  composite time series obtained by joining the indexes of the three different datasets. For what concerns the measurements of H$\alpha$ and NaD, we used only the HARPS and HARPS-N data for which we were able to obtain homogeneous measures (the HIRES H$\alpha$ values cover a different scale).\\
\noindent $\it{HIRES+HARPS+HARPS-N\, -}$ The GLS periodogram for $S$-index (Fig.\, \ref{fig:A4}) indicates a long significant period of $2267 \pm\,50$\,d.\\
\noindent $\it{HARPS+HARPS-N\, -}$ The GLS periodogram for H$\alpha$ indicates a significant period of $2500 \pm\,49$\,d and the residuals time series has a significant period of $40.83 \pm\,0.03$\,d; no significant period was observed for NaD.\\ The period of 40.8 d is the one-year alias of the 37 d rotation period, and the last one is the second significant peak in the H$\alpha$ periodogram shown in Fig.\, \ref{fig:A4}. When removing the long 2500-d period, both the periods gain power, with the alias period growing in power more than the physical rotation period. \\ These periodicities around 2000\,d are apparent also in the whole RV time series in Fig.\,\ref{fig:1}. They indicate that the long-term modulation of the RV is very likely associated with an activity cycle. This period is consistent with those observed for other similar objects (see, e.g., Fig.\,11 in \citealt{sua18}).\\
In the analysis of the activity indicators (when available in the different datasets), no periodic variations around 15.53 d were observed in the $S$-index, the H$\alpha$ and NaD time series, or in the spectral window. \\

\noindent In Fig.\,\ref{fig:5} we show the periodograms of RV data, activity indexes, and spectral window zoomed around the frequency of interest (f=0.064 d$^{-1}$ -- P\,=\,15.53 d), while in Fig.\,\ref{fig:6} we show the 2D representation of the GLS analysis, that is, the periodogram power of the inspected periods of the $S$-index and H$\alpha$ time series as a function of the number of observations. There are several significant periods larger than 100 d, probably due to long trends and a series of significant periods between 29 and 40 d for $S$-index and around 40 d for H$\alpha$.\\ 
The significant periods of the activity indexes, highlighted  by the GLS periodogram analysis, are summarized in Table\,\ref{table:3}, and phase folded plots at those periods are shown in Figs.\, \ref{fig:A5} and \ref{fig:A6}.

\begin{table*}
\centering
\caption{Significant periods (FAP $<$ 1\%) of the activity indexes obtained with the GLS analysis (pre-whitening). The numbers in brackets indicate the data points of the respective time series. }             
\label{table:3}             
\begin{tabular}{ l c c }     
\hline\hline      
              & $S$-index                & H$\alpha$             \\\hline
a HARPS-N (64)& -     & $37.05\,\pm\,0.7\,d$  \\
b HARPS (20)  & - & - \\
c HIRES (114) & $38.00\,\pm\,0.02\,d$    & $45.35\,\pm\,0.05\,d$ \\
a+b (84)      &                 -         & $2500\,\pm\,49\,d$    \\
              &                          & $40.83\,\pm\,0.03\,d$ \\
a+b+c (198)   & $2267\,\pm\,50\,d$       &              -         \\\hline
\hline                 
\end{tabular}
\begin{flushleft}
\end{flushleft}
\end{table*}

\begin{figure*}[!h]
\centering
\includegraphics[width=14cm]{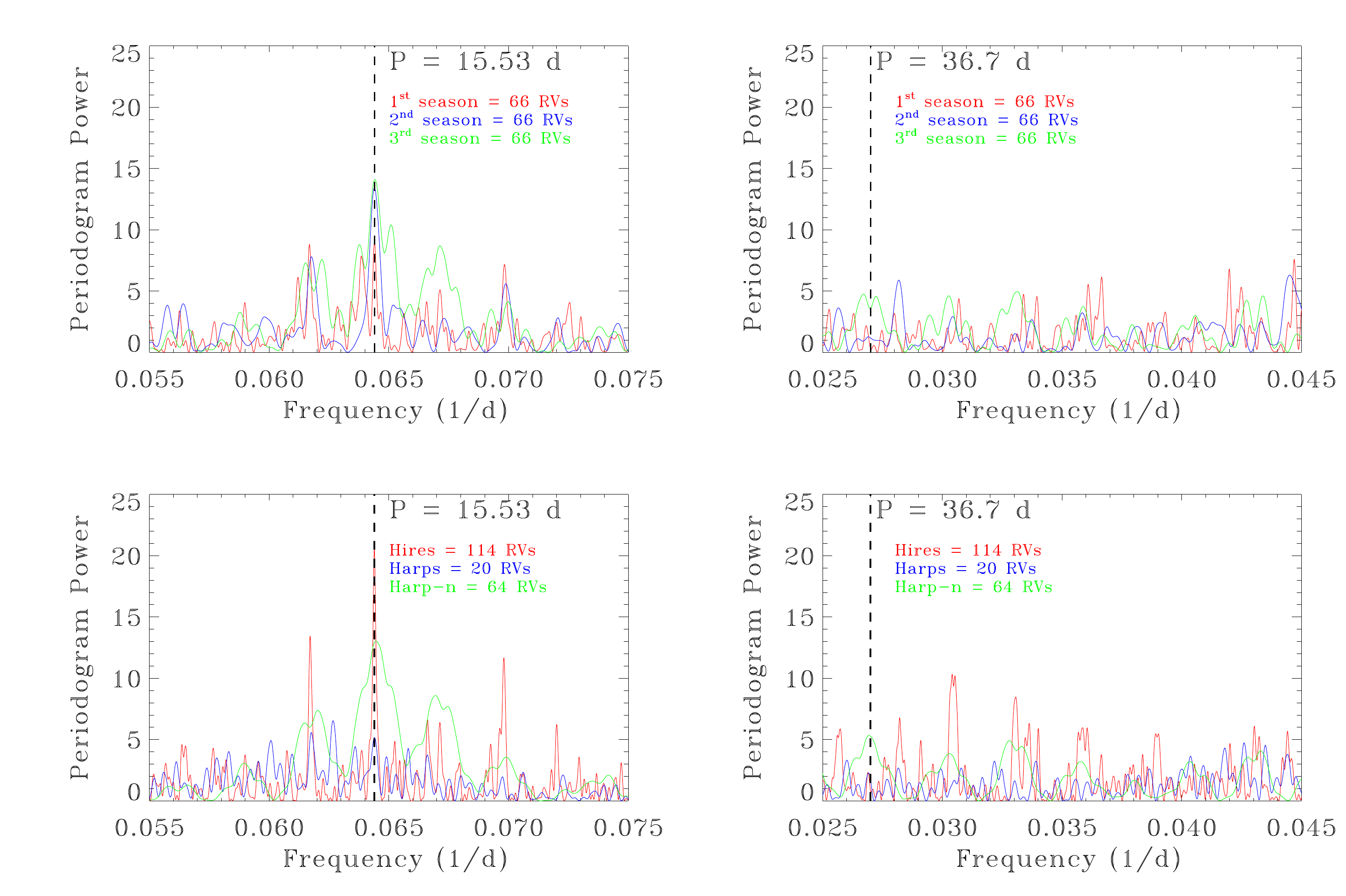}
\caption{GLS periodograms around the frequency ranges of interest. The two upper panels refer to three equal subsets of the composite RV data. Each season has 66 data points: the three seasons, in order, are shown in red, blue, and green, respectively. The two lower panels refer to the separate telescope RV datasets. The HIRES  periodogram is plotted in red, the HARPS in blue, and the HARPS-N in green. The left panels refer to the frequency related to the candidate planet (f\,=\,0.064 d$^{-1}$ -- P\,=\,15.53 d). The right panels refer to the frequency related to activity (f\,=\,0.027 d$^{-1}$ -- P\,=\,36.7 d), as suggested from RV analysis, Gaussian Processes as well as by several activity indicators and ASAS photometry.}
\label{fig:7}
\end{figure*}

To avoid any misinterpretation of the
stellar activity as a planetary signal, we analyzed the combined RV data, subdividing them into three consecutive subsets each containing 66 data points, to check the persistence of the planetary signal over time and to mitigate the possible effects of discontinuities in the data sampling. In the upper plots of Fig.\,\ref{fig:7} we show the GLS periodograms zoomed around the frequency ranges of interest for the three seasons. The red, blue, and green lines show the first, second, and third subset, respectively. In the lower plots we can compare the GLS periodograms for the three telescope datasets, red (HIRES), blue (HARPS), and green (HARPS-N). In the plot we indicate the frequencies obtained from the RV and activity indexes/photometric analysis (see Sect.~\ref{sec:sec4}). The periods related to activity effects span a range from 29 to 45 d (Fig.\,\ref{fig:6}). Significant peaks in the GLS of several activity indicators are not stable during the seasons (37, 38, 40, and 45 d in the three seasons/telescope datasets), and in particular periods of 37 d and 37.8 d are indicated by the Gaussian process (see Sect.~\ref{sec:sec5}) and ASAS photometry (see Sect.~\ref{sec:sec4}), respectively, as estimates of the rotation period. Among the periods identified in the stellar activity indicators, only the period at 36.7 d appears at significant power in RV. We attribute this periodicity to the rotation period of the star. \\The feature at P = 15.53 d (f\,=\,0.064 d$^{-1}$) is observed in the three seasons as the most significant period. The RV amplitudes in the three seasons are: $K_1$ = $2.7\,\pm\, 0.5$\, m $s^{-1}$ (with phase 0.40), $K_2$ = $3.6\,\pm\, 0.5$\, m $s^{-1}$ (with phase 0.40), and $K_3$ = $3.1\,\pm\, 0.4$\, m $s^{-1}$ (with phase 0.54), showing the stability of the signal. We notice that the first season is dominated by HIRES RVs, which are less precise than HARPS/HARPS-N ones, and this is reflected in the RV amplitude $K_1$ and periodogram power, both smaller than in the other two seasons but with the same number of data points. This constitutes further evidence that this signal is present in the RVs at any time (in a time range of more than 20 yrs), leading us to conclude that it has a planetary origin.\\

We computed the correlation between activity indexes and RV time series via the Spearman's rank correlation coefficients. No significant correlation was identified ($\rho$ = -0.01 for $S$-index and $\rho$ = -0.34 for H$\alpha$). We also removed the activity contribution at 36.7 d (adjusting and removing a sinusoidal function with this period) from the RV time series obtaining $\rho$ = -0.01 for $S$-index and $\rho$ = -0.24 for H$\alpha$, thus we do not identify a correlation that would be expected if the RV variation were still induced by activity features on the star surface.\\

    \begin{table*}
    \caption{Best-fit values for the parameters of the global model (one planet on a circular orbit\,+\,stellar activity) selected as explained in Sect. \ref{sec:sec5}. They are calculated as medians of the marginal posterior distributions, and the uncertainties represent their 16$\%$ and 84$\%$ quantiles, unless otherwise indicated. The priors used in our MC fitting are also shown. The difference $\Delta \ln\mathcal{Z}$ between the Bayesian evidence for this model and the one for a `null' model not including a planet is $\Delta \ln\mathcal{Z}=-508.37+530.68=22.31$, which indicates decisive evidence in favor of the one-planet model according to \cite{kass1995}. For the `null' model we  fitted the data only through a GP quasi-periodic regression, using the same priors for the hyper-parameters as for the one-planet model.}    
    \label{table:4}
        \centering
        \begin{tabular}{lll}
        \hline
    \noalign{\smallskip}
    Parameter   &  Value &  Prior \\
    (planet) & & \\
    \noalign{\smallskip}
    \hline
    \noalign{\smallskip}
        $K_{\rm b}$ [$\ms$] &   $3.29^{+0.31}_{-0.32}$  & \textit{$\mathcal{U}$}(0,20) \\ [3pt]
        $P_{\rm b}$  [days]   &  $15.53209^{+0.00166}_{-0.00167}$  &  \textit{$\mathcal{U}$}(13.5, 17.5) \\ [3pt]
        $T_{\rm 0,b}$ [BJD-$2,450,000$] &  $7805.69^{+0.28}_{-0.27}$  &  \textit{$\mathcal{U}$}(7800, 7820)\\ [3pt]
        $e$ & 0 (fixed) & - \\ [3pt]
        \hline 
         & & \\ 
         (hyper)parameter   & Value &  Prior \\
         (stellar activity)  & &   \\ [2pt]
         \hline \\
         $h\, HARPS$ [$\ms$]   &  $1.76^{+0.31}_{-0.28}$ & \textit{$\mathcal{U}$}(0,20) \\ [3pt]
         $h\, HIRES$ [$\ms$]   &  $3.16^{+0.44}_{-0.40}$ & \textit{$\mathcal{U}$}(0,20) \\ [3pt]
         $\lambda$ [days] &   $23.3^{+30.2}_{-18.2}$  & $log \lambda$ \textit{$\mathcal{U}$}(0,3.9) \\ [3pt]
         $w$     &  $0.49^{+0.31}_{-0.18}$  & \textit{$\mathcal{U}$}(0,1) \\ [3pt]
         $\theta$ [days] &  $37.0^{+5.4}_{-14.4}$  & \textit{$\mathcal{U}$}($0, 50$)\\ [3pt]
         $offset_{HARPS-N}$ [\ms] &  $-0.41^{+0.53}_{-0.62}$  & \textit{$\mathcal{U}$}($-100, 100$)\\ [3pt]
         $\sigma_{jitter,HARPS-N}$ [$\ms$] &  $1.44^{+0.29}_{-0.26}$  &  \textit{$\mathcal{U}$}($0, 10$)\\[3pt]
         $offset_{HARPS}$ [\ms] &  $-0.33^{+0.60}_{-0.60}$  & \textit{$\mathcal{U}$}($-100, 100$)\\ [3pt]
         $\sigma_{jitter,HARPS}$ [$\ms$] &  $0.67^{+0.47}_{-0.41}$  &  \textit{$\mathcal{U}$}($0, 10$)\\[3pt]
         $offset_{HIRES}$ [\ms] &  $0.65^{+0.51}_{-0.48}$  & \textit{$\mathcal{U}$}($-100, 100$)\\ [3pt]
         $\sigma_{jitter,HIRES}$ [$\ms$] &  $0.51^{+0.47}_{-0.35}$  &  \textit{$\mathcal{U}$}($0, 10$)\\\noalign{\smallskip}
         \hline
         & & \\
         Derived quantities & & \\
         $M_{\rm b}\sin i$ [$\mearth$] & 7.1$\pm$0.9 & \\
         $a_{\rm b}$ [AU] & 0.091$\pm$0.004 & \\
         $T_{\rm eq, b}$ [K] & 379$^{\rm +24}_{\rm -25}$ & \\
         \noalign{\smallskip}
         \hline
        \end{tabular}
    \tablefoot{The symbol \textit{$\mathcal{U}$}($\cdot$,$\cdot$) denotes an uninformative prior with corresponding lower and upper limits. The equilibrium temperature $T_{\rm eq,b}$ was derived by assuming Bond albedo $A_{\rm B}$=0.}
    \end{table*}

\section{Photometry\\ }\label{sec:sec4}
The targets of the HADES program are photometrically monitored by two independent programs: APACHE and EXORAP (see \citealt{aff16}, for details on instrumental setup and data reduction).
These two programs regularly follow up the sample of M stars to provide an estimate of the stellar rotation periods by detecting periodic modulation in the differential light curves.\\ For \object{Gl\,686} we took advantage also of archive photometric data from the ASAS survey.

\subsection{EXORAP photometry}\label{4.1}
The star \object{Gl\,686} was monitored at INAF-Catania 
Astrophysical Observatory with an 80cm f/8 Ritchey-Chretien robotic telescope (APT2)  located at Serra 
la Nave (+14.973$^{\circ}$E, +37.692$^{\circ}$N, 1725 m a.s.l.) on Mount Etna. We collected $BVR$ photometry of \object{Gl\,686} obtainig 88 data points between 5 May and 18 October 2017. 
We used aperture photometry as implemented in the IDL routine {\em aper.pro}, trying a range of apertures 
to minimize the rms of the ensemble stars. 
To measure the differential photometry, we started with an
ensemble of $\text{about six}$ stars, the  nearest and brightest to \object{Gl\,686}, and checked the variability of each by building their differential light curves 
using the remaining ensemble stars as reference. With this method we selected the least variable stars of the sample, typically four. The photometric variability in the BVR bands is 0.01-0.02 mag, at the level of the night-to-night photometric precision of the survey (Fig.\,\ref{fig:A2}). \\
The differential photometry of the target was analyzed using the GLS algorithm. 
No data were rejected after evaluation of possible outliers (clip at 5$\sigma$). For each band, we analyzed the 
periodograms, assigning them a FAP using the bootstrap method with 10\,000 iterations. The periodograms of the $B$, $V,$ and $R$ magnitudes do not show any peak above the $5\%$ significance level.

\subsection{APACHE photometry}\label{4.2}
    APACHE is a photometric survey devised to detect transiting planets around hundreds of early-to-mid-M dwarfs \citep{soz13}.
     \object{Gl\,686} was monitored for 49 nights between March 29, 2015, and July 1, 2016, with one of the five 40-cm telescopes composing the APACHE array, located at the Astronomical Observatory of 
     the Autonomous Region of the Aosta Valley (OAVdA, +45.7895 N, +7.478 E, 1650 m.a.s.l.).\\
The APACHE data were obtained using an aperture of 4.5 pixels and a set of four comparison stars. 
We searched the light curve for sinusoidal-like modulation by using the complete dataset consisting of 555 points, each being the average of three consecutive measurements with an uncertainty equal 
to their rms. The APACHE data show a long-term trend. We used the GLS algorithm to calculate the frequency periodograms. To estimate the significance of the detection, we performed a bootstrap analysis (with replacement) using 10\,000 permuted datasets derived from the original photometric data.
The best photometric frequency we find is associated to a period of 29 d in the original dataset and 27 d in the detrended dataset (with a FAP $\le$\,10\%), both of them are close to those found in the RVs and spectroscopic activity index time series, whose power is nonsignificant.

\subsection{ASAS photometry}\label{4.3}
We also analyzed photometry of \object{Gl\,686} from the All-Sky Automated Survey \citep[ASAS][]{poj97}, which overlaps with HIRES and HARPS data, from March 3, 2003, and May 17, 2009, finding evidence of the stellar rotation period at $P_{rot}$\,=\,37.79 $\pm$\,0.08 d. This value is compatible with the stellar rotation period also found in the HARPS-N RV data (36.7 d) and in the activity indicators (37 d in H$\alpha$ and 38 d in S index).\\
We used an autocorrelation analysis \citep{box76,ede88}, only to confirm the periodicity of the
ASAS light curve. The time series
is random if it consists of a series of independent observations with the same distribution. In this case we would expect the autocorrelation of the time series to be statistically nonsignificant for all values of the lag. The ASAS light curve has an autocorrelation coefficient relative to a period of 36 d significant at the 2$\sigma$ level (i.e., the periodogram and autocorrelation analysis agree within the error bars).

\section{Monte Carlo analysis of the RV time series based on Gaussian process regression}
        \label{sec:sec5}
To mitigate the stellar activity signal in the RV time series, and to test the statistical significance of a dynamical model including one planet over the `null hypothesis' case, where only an activity term is considered, we have defined a Bayesian framework based on a Monte Carlo (MC) sampling of the parameter space. The stellar activity term has been treated as a quasi-periodic signal through a Gaussian process (GP) regression, a technique commonly used to detect or characterize many exoplanetary systems \citep{hay14}, also within the HADES collaboration \citep{aff16,per17b,pin18}. We used a quasi-periodic kernel to model the correlated RV signal due to stellar activity, and we refer to one of these latter papers for details of the quasi-periodic model. Here we summarize the meaning of the GP hyper-parameters included in the quasi-periodic kernel, as discussed in this work: $h$ is the amplitude of the correlations (we consider two $h$ parameters, one in common for HARPS and HARPS-N and one for HIRES, on account of possible differences related to the use of different instruments); $\theta$ is the recurrence time of the correlations (generally, the stellar rotation period); the parameter $w$ describes the level of high-frequency variation within a single stellar rotation, and it is related to the distribution of the features on the stellar surface related to the magnetic activity; and $\lambda$ is the correlation decay time scale, which can be physically related to the lifetime of the magnetically active regions.

\begin{figure*}
\centering
\includegraphics[width=18cm]{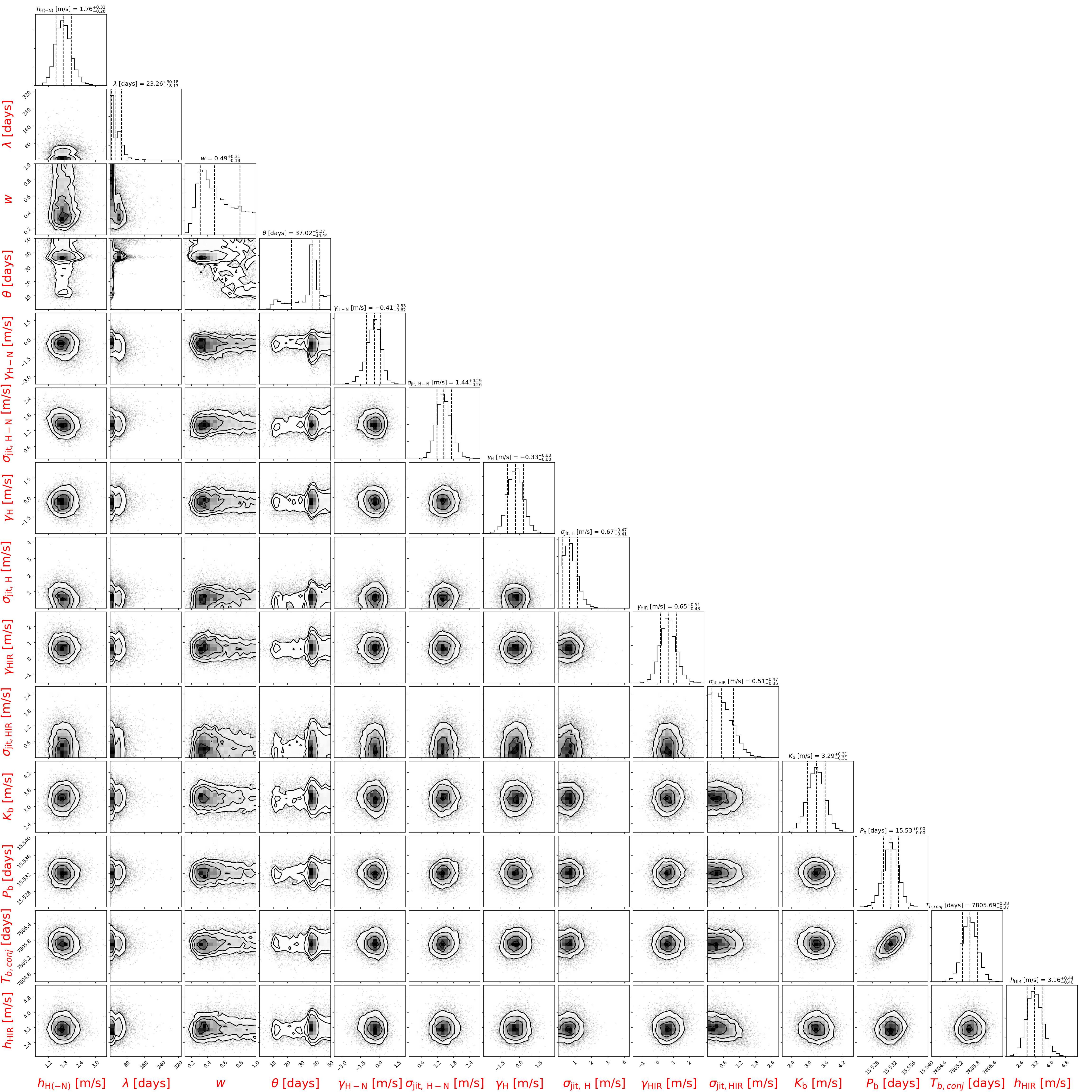}
\caption{Marginal posterior distributions for the parameters of our model, which includes one Keplerian, for the RV data. H stands for HARPS, H-N for HARPS-N, and HIR for HIRES, in the labels.}
\label{fig:8}
\end{figure*}

The use of the quasi-periodic GP regression in our work seems justified in view of the previously described analysis, which showed that signals modulated at the stellar rotation have been identified in ancillary data and could also
be present in the RV dataset. We have first fitted the RVs without including a Keplerian in the model, which then represents the `null scenario' against which the model with one planet can be compared within a Bayesian framework. We performed the MC sampling within a Gaussian process framework using the open-source Bayesian inference tool \texttt{MultiNestv3.10} (e.g. \citealt{fer13}), through the \texttt{python} wrapper \texttt{pyMultiNest} \citep{buc14}, with 800 live points and a sampling efficiency of 0.3. We integrated into our code the publicly available GP module \texttt{GEORGEv0.2.1} \citep{amb14}. Table \ref{table:4} summarizes the best-fit values for the (hyper-)parameters of the one-planet circular model, which includes RV offsets and uncorrelated jitter terms for each instrument. \textcolor[rgb]{0.984314,0.00784314,0.027451}{\textcolor[rgb]{0,0,0}{This latter model}} is very strongly favored over the `null scenario' and even more strongly favored over the one-planet eccentric model, according to the Bayesian evidence ($\ln\mathcal{Z}$) calculated by \texttt{MultiNest} ($\Delta \ln\mathcal{Z}=\ln\mathcal{Z}_{\rm 1-plan,\: circ}-\ln\mathcal{Z}_{\rm other model}$=22 and 25, respectively). In all cases we have used a large uninformative prior on the hyper-parameter $\theta$, and the posterior for the case of the one-planet model is quasi symmetric and narrowly centered around $\sim$37 days. This result is in very good agreement with the expected stellar rotation period, as derived from the analysis of ancillary data. As a consequence of our choice of relatively unconstrained priors for the stellar activity term, the hyper-parameter $\lambda$ shows a bi-modal posterior, with the maximum a posteriori (MAP) value of $\sim$ 44 days. The first mode with the higher values corresponds to samples with higher $\theta$: in fact, by selecting only samples for which $\lambda$ > 16 days, we get $\theta$=38$^{\rm +4}_{\rm -2}$ days. This is to be expected, since a higher evolutionary timescale of the active regions should imply a more precise determination of the stellar rotation period. However, we note that the planetary parameters are not affected by the particular representation of the stellar activity term. Our unconstrained GP analysis shows that a signal modulated on the expected stellar rotation period is indeed present in the data. In Fig. \ref{fig:8} we show the marginal posterior distributions for the parameters of our model, one planet on a circular orbit + stellar activity.\\
We tested the nature of the 15.5-day signal by applying the apodized Keplerian model proposed by \cite{gre16}, in which the semi-amplitude $K$ of the planetary orbital equation is multiplied by a
symmetrical Gaussian of unknown width $\tau$ and with an unknown
center of the apodizing window $t_{\rm a}$. We obtain clear evidence that $\tau$ converges toward values much higher than the time span of the data ($\tau$=20278$^{\rm -7827}_{\rm -8250}$ days, with prior $\mathcal{U}$(1,32\,000)), indicating that $K$ is constant over the range of the observations. \\ As a second test on the nature of the 15.5-d signal, we performed a GP + 1 Keplerian planetary signal model, using three different inferior conjunction times (T$_0$), one for each season described in Sec. \ref{3.4}, but leaving the set-up of the general parameters substantially unmodified. Each T$_0$ has been searched around a value close to the center of each observing season, in order to obtain similar uncertainties for each of them. Using the best-fit orbital period, the T$_0$ posterior distribution has been scaled to a single value, obtaining as a result [BJD-2450000]: T$_{01}$\,=$5615.15^{+ 0.48}_{-0.46}$ days, T$_{02}$\,=$5615.4\,\pm\, 0.36$ days, and T$_{03}$\,=$5616.31^{+ 0.35}_{-0.33}$ days.
The T$_0$ values of the first two seasons are well within the errors (0.55 $\sigma$), while T$_{03}$ is 1.6 $\sigma$\, from T$_{02}$. Therefore, these two tests support the planetary interpretation of the RV signal.

\begin{figure*}
\centering
\includegraphics[width=11cm]{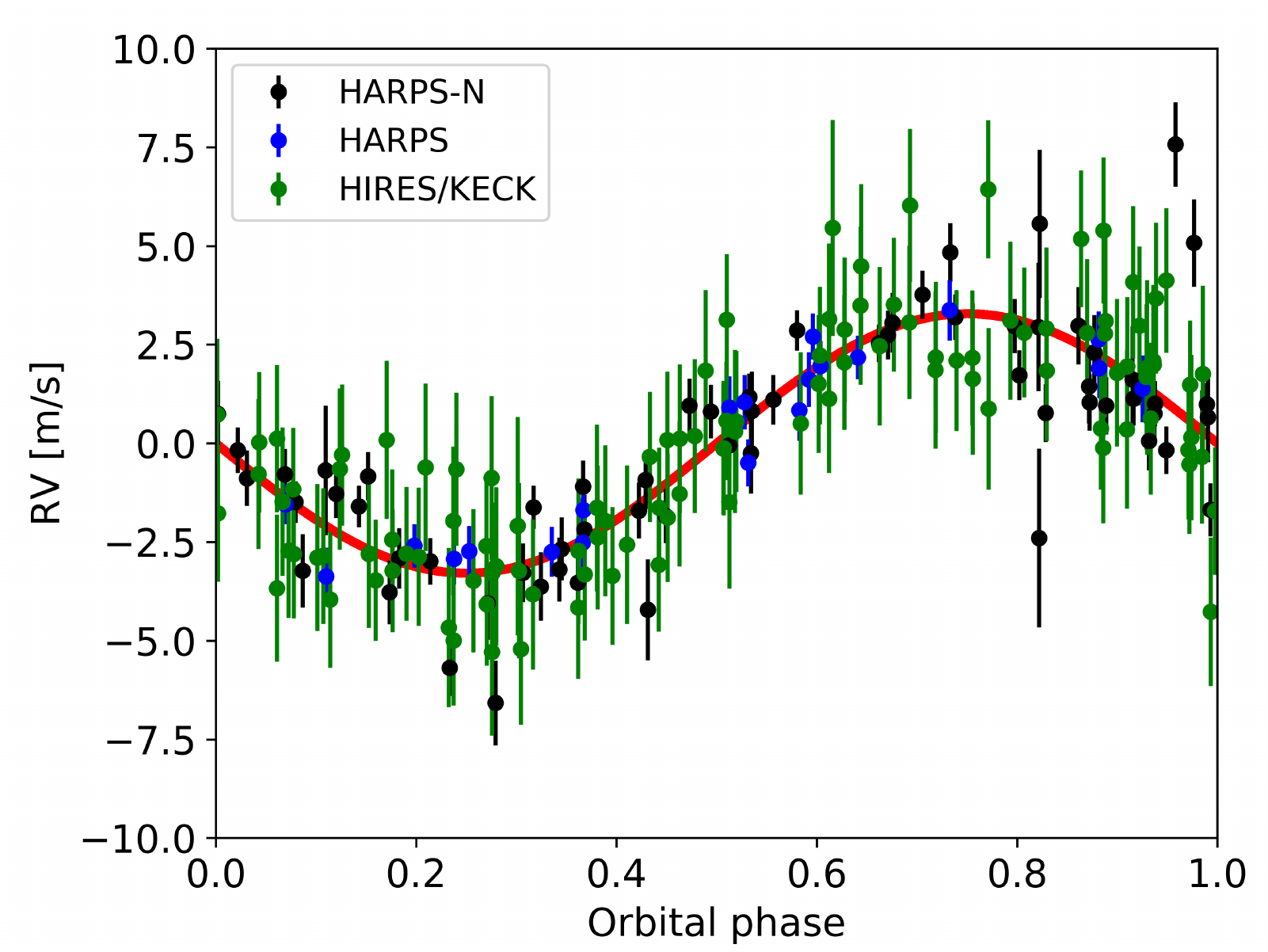}
\caption{RV data folded at the best-fit orbital period of the planet. The red solid line is the best-fit orbital solution. }
\label{fig:9}
\end{figure*}

\begin{figure*}
\centering
\includegraphics[width=14cm]{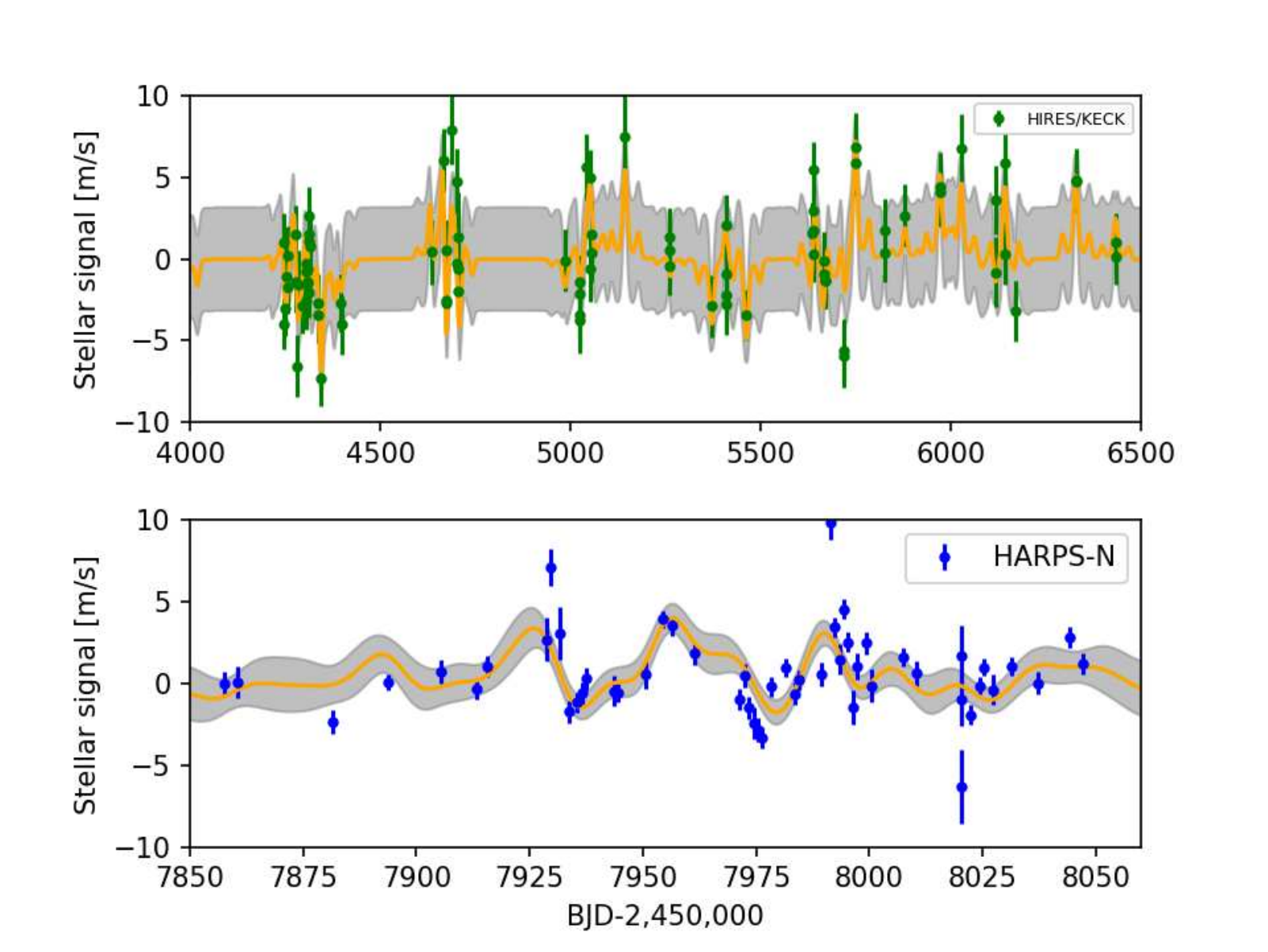}
\caption{Blow-up of the best-fit quasi-periodic signal of two particularly crowded regions (orange line) for HIRES data (upper panel) and HARPS-N (lower panel), the grey areas mark the uncertainties band $\pm$\,1 $\sigma$.}
\label{fig:10}
\end{figure*}

We used the results of the one-planet circular model to verify whether or not the planetary signal has a significant eccentricity by imposing a narrow uniform prior on $\theta$ ($\mathcal{U}(34,42)$\,days) and a larger prior on $P_{\rm b}$ ($\mathcal{U}(0,20)$\,days). We recover a solution which is very similar to that for the circular case, where the eccentricity is compatible with zero ($e$=0.05$^{\rm +0.05}_{-0.03}$). Figure \ref{fig:9} shows the RV curve folded at the best-fit orbital period for the detected planet. In Fig. \ref{fig:10} we show the RV residuals after the Keplerian has been removed, with the best-fit stellar, correlated, quasi-periodic signal superposed.\\
We also tested a model involving two Keplerian signals which, as expected from our previous analyses, appears less probable than that with one planet ($\ln\mathcal{Z}=-510.63\,\pm\,0.27$). With this model we always obtain the stellar-rotation period, as before, and it is worth noticing that the parameters of the planet \object{Gl\,686\,b}\, remain almost unchanged in the transition from models with a different number of planets.

    \begin{table*}
    \caption{Best fit values for the hyper parameters of the covariance function, derived from a GP analysis of the H$\alpha$ activity index time series.}    
    \label{table:5}
        \centering
        \begin{tabular}{lll}
        \hline
    \noalign{\smallskip}
    Parameter   &  Value &  Prior \\
    \noalign{\smallskip}
    \hline
    \noalign{\smallskip}
         $h$    &  $0.01061^{+0.00255}_{-0.00192}$ & \textit{$\mathcal{U}$}(0,20) \\ [3pt]
         $\lambda$ [days] &   $251^{+152}_{-126}$  & \textit{$\mathcal{U}$}(0,8000) \\ [3pt]
         $w$     &  $0.872^{+0.091}_{-0.149}$  & \textit{$\mathcal{U}$}(0,1) \\ [3pt]
         $\theta$ [days] &  $39.38^{+2.32}_{-2.28}$  & \textit{$\mathcal{U}$}($0, 50$)\\ [3pt]
         \hline
        \end{tabular}
    \end{table*}

\subsection{Bayesian analysis of the H$\alpha$ time series}\label{5.1}
We performed the Bayesian analysis of the H$\alpha$ index, obtained from the HARPS and HARPS-N spectra, corrected for instrumental offset. We used the same prior as in the case of RV analysis (see Table \ref{table:5}). \\
The H$\alpha$ analysis indicates that the distribution obtained for $\theta$ is bimodal (1-year alias) and the first peak is around 37 d, which is the same result we obtained in the planetary case. The $\lambda$ value is larger than that obtained considering the RVs time series, though the two datasets are not comparable (H$\alpha$ data are only from the two HARPS instruments), both in terms of time span and number of data points.  \\

\begin{figure*}
\includegraphics[width=14.cm]{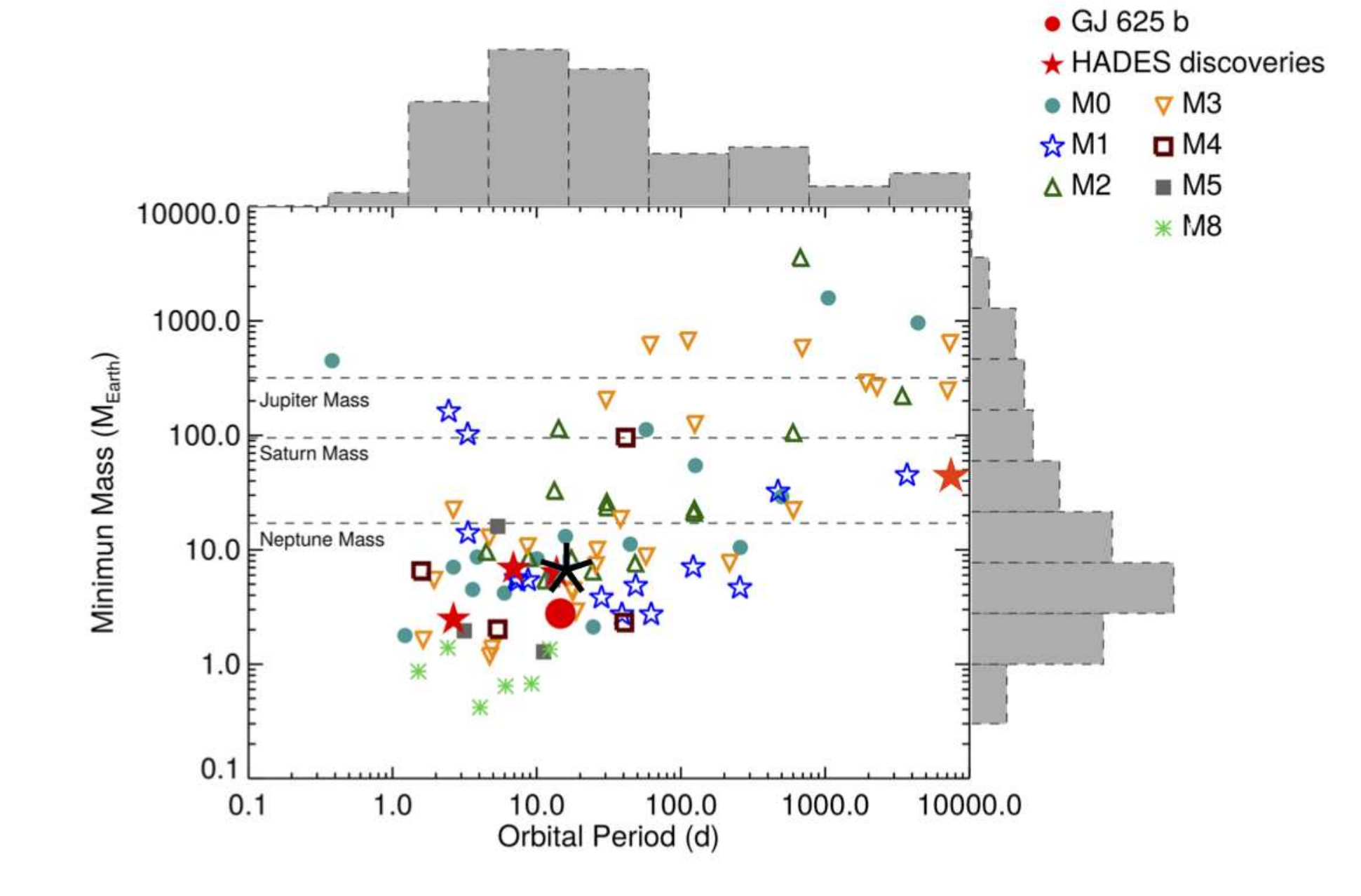}
\caption{Adaptation of Fig. 21 from \citet{sua17}, showing minimum mass vs. orbital period for the known planets with determined masses around M-dwarf stars (using exoplanet.eu data) divided by the spectral type of the host star. The red filled stars show the positions of HADES planets, the red filled dot shows the position of the planet \object{GJ\,625\,b}, the black asterisk has been added to show the position of \object{Gl\,686\,b}. The horizontal dashed lines show the mass of the solar-system planets for comparison. The distribution for each parameter is shown at the edges of the figure.}
\label{fig:11}
\end{figure*}

\section{DISCUSSION}\label{sec:sec6} 
Several works on M stars \citep[e.g.,][]{kur03,rob14,rob15} have highlighted that the activity signals that create RV shifts might not create a detectable photometric periodicity. This can happen in several cases when magnetic phenomena inhibit the convection but do not create dark spots on the surface of the star, thereby altering RVs without a corresponding photometric signature. We observed this phenomenon also for \object{Gl\,686}, for which the ancillary photometric observations, but for ASAS photometry, do not detect a clear periodicity.\\
We can speculate about the physical meaning of the recurrent period at 45 d (both in RV and activity time series), ascribing it to differential rotation, that is, latitudinal differences in the rotation velocity. In general, differential rotation is not yet fully understood, as is discussed in  \citet{kuk11}. In their work on solar-type stars, \citet{rei03} considered a calibration of the ratio between the positions of the two first  zeros of the power spectrum of the line profile (in the Fourier domain) as a function of the ratio $\alpha$ between equatorial and polar rotation rate. They obtained an anticorrelation between differential rotation and equatorial velocity: differential rotation is mostly evident in slow rotators, such as \object{Gl\,686}. However, Reiners \& Schmitt only considered solar-type stars and did not show that we might see such trends in time series of M dwarfs.
\citet{rei15} analyzed an extremely wide dataset of stars observed with Kepler, ranging from solar-type stars to M dwarfs, and confirmed the relation found by Reiners \&  Schmitt between differential rotation and period. We may directly compare our result for \object{Gl\,686} with Fig. 9 of Reinhold \& Gizon, where they plotted the minimum rotation period (in our case 37 d),
with the parameter $\alpha$=(Pmax-Pmin)/Pmax. For \object{Gl\,686}, $\alpha$=(45-37)/45=0.177. The point for \object{Gl\,686} falls exactly in the middle
of the points for the M stars. We conclude that the assumption that these two periods related to activity for \object{Gl\,686} are due to rotation is fully consistent with what we know about differential rotation. \\
Looking at the $\it Gaia$ solution, it appears that the residues in the astrometric solution are small and there is no evidence of significant noise. This point, coupled with the fact that there is no trend in RV, suggests that there are no giant planets in the system, even at separations of some astronomical units. Moreover, $\it Gaia$ did not find any companion within a degree; this also places limits on massive brown dwarfs with great separation (a result that depends on age, which is unknown for \object{Gl\,686}).
The results within the HADES survey so far confirm that the frequency of giant companions around M stars is lower than that around solar-type hosts, a result that is now rather well observationally established, as clearly shown in \citet{end06} \citep[see also,][and references therein]{soz14,dre15}.\\
The main interest in focusing the research of extrasolar planets around M stars is to discover an Earth twin, a planet lying in the HZ of the system star-planet. However, the topic is very complex and apart from the star-planet separation and stellar luminosity, many other factors also influence habitability, such as planetary environment (e.g., planet albedo), composition, and the amount of greenhouse gases in the atmosphere \citep{kal10}. Moreover, planets in close orbits around low-mass stars are probably tidally locked, thus only one side of the planet is strongly irradiated by the star and the possible redistribution of the incident energy on both hemispheres is strictly subjected to the existence of an atmosphere, to its composition, and to circulation. \\The equilibrium temperature of the planet is derived using the Stefan-Boltzmann law (ignoring the greenhouse effect):
\begin{equation}
 T_{eq} = T_{\star}{(\frac{R_{\star}}{2a})}^{1/2}[f(1-A_{B})]^{1/4},
\end{equation}
where $T_{\star}$ is the stellar effective temperature,  $a$ is the semi-major axis of the planet, $A_{B}$ is the Bond albedo, and the $f$ factor describes the atmospheric circulation. If the planet is tidally locked to its star, but advection distributes the incident energy over both hemispheres, we can assume $f$\,=\,1. If only the day side reradiates the incident energy, the higher equilibrium temperature is given by $f$\,=\,2 \citep[e.g.,][]{cha05,men17}. Therefore, we can derive an upper limit on $ T_{eq}$ by setting $A_{B}$\,=\,0. In the two extreme cases, $f$\,=\,1 and $f$\,=\,2, for \object{Gl\,686\,b}\, $T_{eq}$\,=\,379 K and 450 K, respectively.\\ We estimated the conservative inner limits of the habitable zone for \object{Gl\,686}, following \citet{kas93} and \citet{kop13}. The optimistic inner limit ($\it recent\, Venus$) corresponds to a semi-major axis of 0.137 AU, for the HZ of \object{Gl\,686}. To estimate a more optimistic inner limit, we followed the empirical fitting formulas of \citet{zso13}, which provide an inner edge limit of 0.1 AU, for a surface albedo of 0.2 and 1\% relative humidity, and a limit of 0.07 AU, for a surface albedo of 0.8. Thus, we can tentatively state that \object{Gl\,686\,b}\, could be located inside, or near the inner limit of the HZ, depending on its atmospheric conditions.\\

\section{Summary and conclusions}\label{sec:sec7}
For the star \object{Gl\,686}, we took advantage of our own observing campaign within the HADES program with HARPS-N, and of public archive data, obtained with HARPS and HIRES. The composite dataset covers a time span of more than 20 years. 
The very wide coverage of RV data and the persistence of the signal over such a long time span make \object{Gl\,686} an interesting target not only for planet detection purposes but, particularly, for the study of the behavior of activity in M dwarfs. \\We detected one significant RV signal at a period of 15.53 d, both in each one of the datasets and in the composite time series. The dominant periodicity shows a very high spectral power. In order to shed light on the real origin of the signal we performed a complete analysis of the RV and activity indexes time series as well as a study of the photometric light curve, using the Gaussian processes amongst other techniques. The periods related to activity effects span a range from 27 to 45 d, and the significant peaks in the GLS of several activity indicators (related to magnetic phenomena such as spots and faculae, which rotate on the stellar surface, and to differential rotation) are not stable during the seasons or datasets (37, 38, 40 and 45 d in the three seasons/telescope datasets). The second RV signal of a period of 36.7 d in HARPS-N data, also suggested by several activity indicators (H$\alpha$ 37.05 d in HARPS-N data and $S$-index 38 d in HIRES data) and by the Gaussian Process (37 d) and ASAS photometry (37.8 d), provides an estimate of the rotation period and also of the variability of the spot configuration during the long time span, and the hint of an activity cycle $\sim$\,2000 d long. Conversely, the spectral feature at 15.53 d is a coherent signal, observed to stay stable during all three seasons.\\
We used a Bayesian model selection to determine the probability of the existence of a planet in the system and to estimate the orbital parameters and minimum mass. We tested several different models in which we varied the number of planets (0, 1, 2) and the eccentricity, varying the prior ranges. We selected a model involving one Keplerian circular signal and a GP describing stellar activity variations (the two Keplerian + GP model has a lower Bayesian evidence). \\From the available photometric and spectroscopic information we conclude that the 15.53-d signal is caused by a planet with a minimum mass of $7.1\pm\, 0.9$\, $\mearth$, in an orbit with a semi-major axis of 0.09 AU.\\
\object{Gl\,686\,b}\, goes to populate the lower part of the diagram of minimum mass versus orbital period of the known planets around M stars, as well as the other super-Earths discovered within the HADES program, \object{GJ\,3998\,b}\, ,\object{GJ\,3998\,c} \citep{aff16}, \object{GJ\,625\,b} \citep{sua17}, and \object{GJ\,3942\,b} \citep{per17b} (see Fig. \ref{fig:11}, adapted from \citealt{sua17}). \\Following the preliminary results of the HADES survey, one characteristic of the behavior of M dwarf planetary systems that seems to emerge is the recurrence of planetary systems with planet minimum masses lower than 10 $\mearth$, orbital periods in the range 10-20 d, and stellar rotation periods between 30 and 40 d.

\begin{acknowledgements}
GAPS acknowledges support from INAF through the Progetti Premiali funding scheme of the Italian Ministry of Education, University, and Research. LA, GM, JM, AG and MD acknowledge financial support from Progetto Premiale 2015 FRONTIERA (OB.FU. 1.05.06.11) funding scheme of the Italian Ministry of Education, University, and Research.
GS acknowledges financial support from ``Accordo ASI-INAF'' No. 2013-016-R.0 July 9, 2013 and July 9, 2015. MP, and IR, acknowledge support from the Spanish Ministry of Economy and Competitiveness (MINECO) and the Fondo Europeo de Desarrollo Regional (FEDER) through grant ESP2016-80435-C2-1-R, as well as the support of the Generalitat de Catalunya/CERCA program.
JIGH, RR and BTP acknowledge financial support from the Spanish Ministry project MINECO AYA2017-86389-P, and JIGH from the Spanish MINECO under the 2013 Ram{\'o}n y Cajal program MINECO RYC-2013-14875. MP gratefully ackowledges the support from the European Union Seventh Framework program (FP7/2007-2013) under Grant Agreement No. 313014 (ETAEARTH).\\ This work is based on observations made with the Italian Telescopio Nazionale Galileo (TNG), operated on the island of La Palma by the Fundaci{\'o}n Galielo Galilei of the Istituto Nazionale di Astrofisica (INAF) at the Spanish Observatorio del Roque de los Muchachos (ORM) of the Instituto de Astrof{\'{i}}sica de Canarias (IAC).\\
We wish to thank the anonymous referee for thoughtful comments and
suggestions, which helped to improve the manuscript.
\end{acknowledgements}

\bibliographystyle{aa}
\bibliography{Bibliographyrv}

\appendix
\section{Appendix}
In this section we report the observational data collected with the HARPS-N and HARPS spectrographs (from TERRA pipeline), the uncertainties are the internal errors; the time series plots of the activity indexes ($S$-index, H$\alpha$, NaD); the time series plots of the ASAS and EXORAP photometry; the phase folded plots, at the most significant periods of the GLS analysis of activity indexes.
 \onecolumn
\begin{center}
\begin{longtable}{c|c|c|c }
\caption{\label{TabA.1} RV time series from HARPS and HARPS-N.   }\\
 
  \hline \hline
 BJD  & RV    & dRV   & Instrument \\
 (d)    & (m $s^{-1}$)   & (m $s^{-1}$)   & \\ \hline 
 \endfirsthead
 \caption{continued.}\\
 
 \hline \hline
 BJD  & RV    & dRV   & Instrument \\
 (d)    & (m $s^{-1}$)   & (m $s^{-1}$)   & \\ \hline 
 \endhead
\hline
 \endfoot
   2453159.74925 & 3.622 &0.708 & HARPS        \\
   2453574.62253 & 0.683 &0.694 & HARPS        \\
   2453817.86816 &-1.087 &0.640 & HARPS        \\
   2454174.87132 &-1.876 &0.646 & HARPS        \\
   2454194.90763 & 2.401 &0.691 & HARPS        \\
   2454300.64483 &-2.806 &0.631 & HARPS        \\
   2454948.86153 &-1.510 &0.506 & HARPS        \\
   2454950.86206 &-2.799 &0.545 & HARPS        \\
   2454956.83070 &-0.860 &0.770 & HARPS        \\
   2455390.64094 & 0.913 &0.793 & HARPS        \\
   2455392.63268 & 1.920 &0.551 & HARPS        \\
   2455407.58589 & 0.750 &0.630 & HARPS        \\
   2455409.58801 & 2.104 &0.765 & HARPS        \\
   2455412.57851 & 0.209 &0.845 & HARPS        \\
   2455437.52933 &-3.101 &0.592 & HARPS        \\
   2455438.53350 & 0.0   &0.583 & HARPS        \\
   2455446.52185 &-6.482 &0.725 & HARPS        \\
   2455450.49607 &-4.481 &1.137 & HARPS        \\
   2455450.50812 &-3.653 &0.795 & HARPS        \\
   2455458.50534 & 0.540 &0.750 & HARPS        \\
   2456700.74950 & 3.449 &0.976 &  HARPS-N     \\
   2456702.75851 & 1.131 &1.143 &  HARPS-N     \\
   2457508.59296 &-0.499 &0.720 &  HARPS-N     \\
   2457510.60844 &-0.557 &0.841 &  HARPS-N     \\
   2457536.60127 & 0.890 &0.755 &  HARPS-N     \\
   2457537.57229 & 0.927 &0.576 &  HARPS-N     \\
   2457538.56821 &-0.596 &0.630 &  HARPS-N     \\
   2457606.46289 &-7.009 &0.809 &  HARPS-N     \\
   2457608.52949 &-6.822 &0.744 &  HARPS-N     \\
   2457609.48134 &-5.751 &0.881 &  HARPS-N     \\
   2457610.46819 &-7.729 &1.278 &  HARPS-N     \\
   2457857.64182 &-3.148 &0.800 &  HARPS-N     \\
   2457860.58996 & 0.398 &1.000 &  HARPS-N     \\
   2457881.61942 &-0.658 &0.725 &  HARPS-N     \\
   2457893.62668 & 2.509 &0.480 &  HARPS-N     \\
   2457905.54171 &-1.052 &0.734 &  HARPS-N     \\
   2457913.42783 & 0.568 &0.659 &  HARPS-N     \\
   2457915.48199 &-0.733 &0.643 &  HARPS-N     \\
   2457928.62545 & 3.966 &1.357 &  HARPS-N     \\
   2457929.57855 & 7.202 &1.111 &  HARPS-N     \\
   2457931.64334 & 0.590 &1.638 &  HARPS-N     \\
   2457933.57005 &-5.395 &0.713 &  HARPS-N     \\
   2457935.55273 &-4.075 &0.632 &  HARPS-N     \\
   2457936.50234 &-2.522 &0.708 &  HARPS-N     \\
   2457937.61603 &-0.228 &0.677 &  HARPS-N     \\
   2457943.56272 & 1.447 &0.943 &  HARPS-N     \\
   2457944.50282 & 0.271 &0.568 &  HARPS-N     \\
   2457950.51349 &-2.784 &0.864 &  HARPS-N     \\
   2457954.48680 & 5.138 &0.510 &  HARPS-N     \\
   2457956.43367 & 6.301 &0.611 &  HARPS-N     \\
   2457961.48733 & 0.790 &0.701 &  HARPS-N     \\
   2457971.42438 & 1.530 &0.621 &  HARPS-N     \\
   2457972.39543 & 3.350 &0.744 &  HARPS-N     \\
   2457973.39354 & 1.255 &0.687 &  HARPS-N     \\
   2457974.54434 &-0.441 &0.942 &  HARPS-N     \\
   2457975.47540 &-1.869 &0.741 &  HARPS-N     \\
   2457976.42985 &-3.568 &0.670 &  HARPS-N     \\
   2457978.40128 &-2.844 &0.608 &  HARPS-N     \\
   2457981.46533 &-2.406 &0.545 &  HARPS-N     \\
   2457983.51089 &-2.040 &0.686 &  HARPS-N     \\
   2457984.49862 & 0.076 &0.592 &  HARPS-N     \\
   2457989.40811 & 3.048 &0.732 &  HARPS-N     \\
   2457991.41740 &10.348 &1.070 &  HARPS-N     \\
   2457992.41040 & 2.595 &0.575 &  HARPS-N     \\
   2457993.41983 &-0.627 &0.927 &  HARPS-N     \\
   2457994.43299 & 1.466 &0.616 &  HARPS-N     \\
   2457995.39168 &-1.058 &0.592 &  HARPS-N     \\
   2457996.40619 &-5.063 &1.074 &  HARPS-N     \\
   2457997.39394 &-2.061 &0.814 &  HARPS-N     \\
   2457999.41286 & 1.526 &0.688 &  HARPS-N     \\
   2458000.36403 & 0.180 &1.018 &  HARPS-N     \\
   2458007.43803 & 1.455 &0.547 &  HARPS-N     \\
   2458010.44627 &-2.780 &0.767 &  HARPS-N     \\
   2458020.35765 & 1.627 &1.628 &  HARPS-N     \\
   2458020.36872 &-3.716 &2.265 &  HARPS-N     \\
   2458020.37951 & 4.244 &1.877 &  HARPS-N     \\
   2458022.34095 &-1.270 &0.603 &  HARPS-N     \\
   2458024.36957 &-2.123 &0.528 &  HARPS-N     \\
   2458025.35424 &-1.982 &0.528 &  HARPS-N     \\
   2458027.34820 &-4.022 &0.935 &  HARPS-N     \\
   2458031.40518 & 1.328 &0.593 &  HARPS-N     \\
   2458037.36400 & 1.270 &0.675 &  HARPS-N     \\
   2458044.36215 & 0.0   &0.665 &  HARPS-N     \\
   2458047.31078 & 1.944 &0.626 &  HARPS-N     \\ 
\hline \hline
\end{longtable}
\end{center}

\begin{figure*}
\centering
\includegraphics[width=14cm]{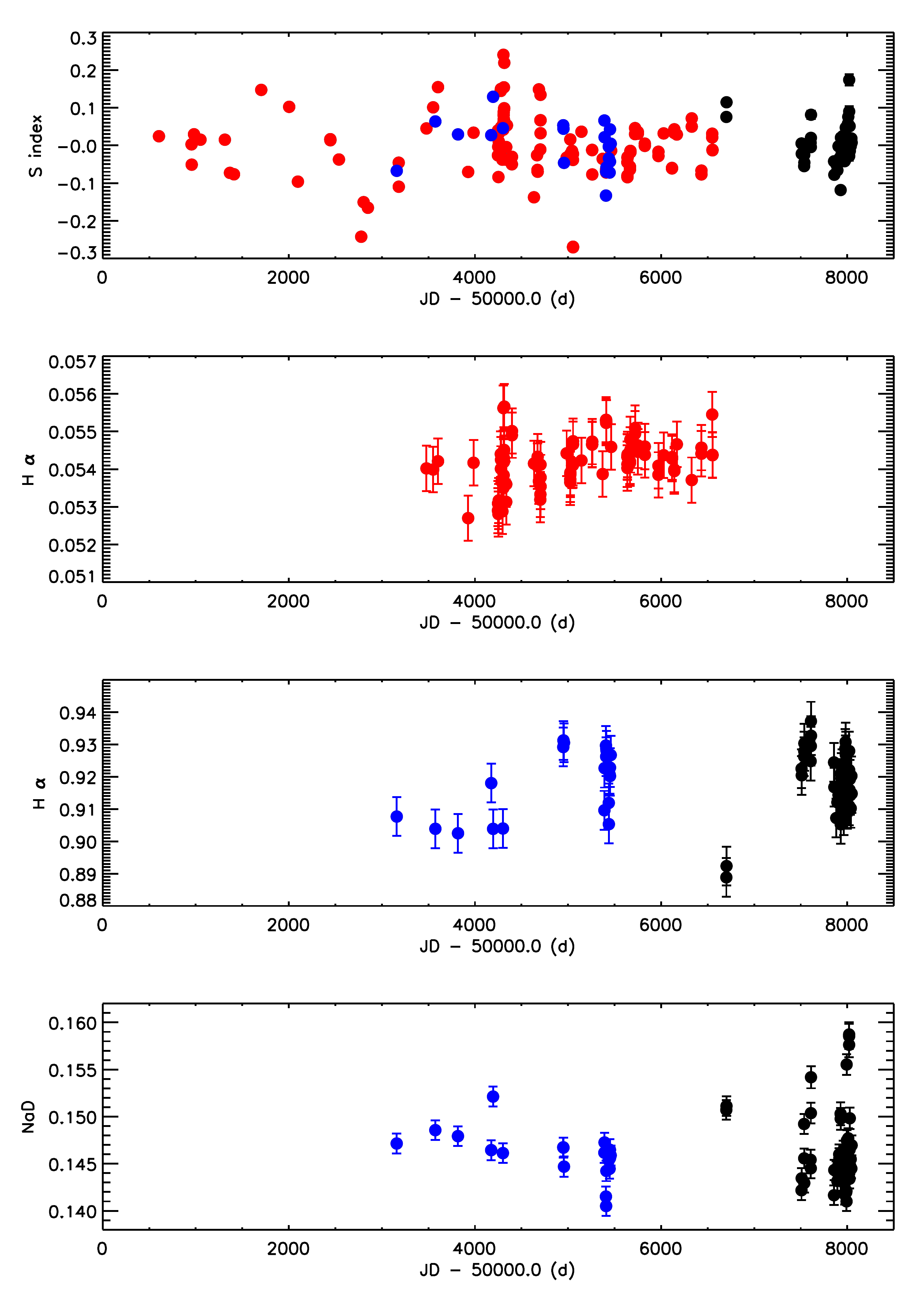}
\caption{Time series data of the activity indexes, $S$-index (HIRES/red, HARPS/blue, HARPS-N/black), H$\alpha$ (HIRES/red), H$\alpha$, and NaD (HARPS/blue, HARPS-N/black), from top to bottom.}
\label{fig:A1}
\end{figure*}

\begin{figure*}
\centering
\includegraphics[width=14cm]{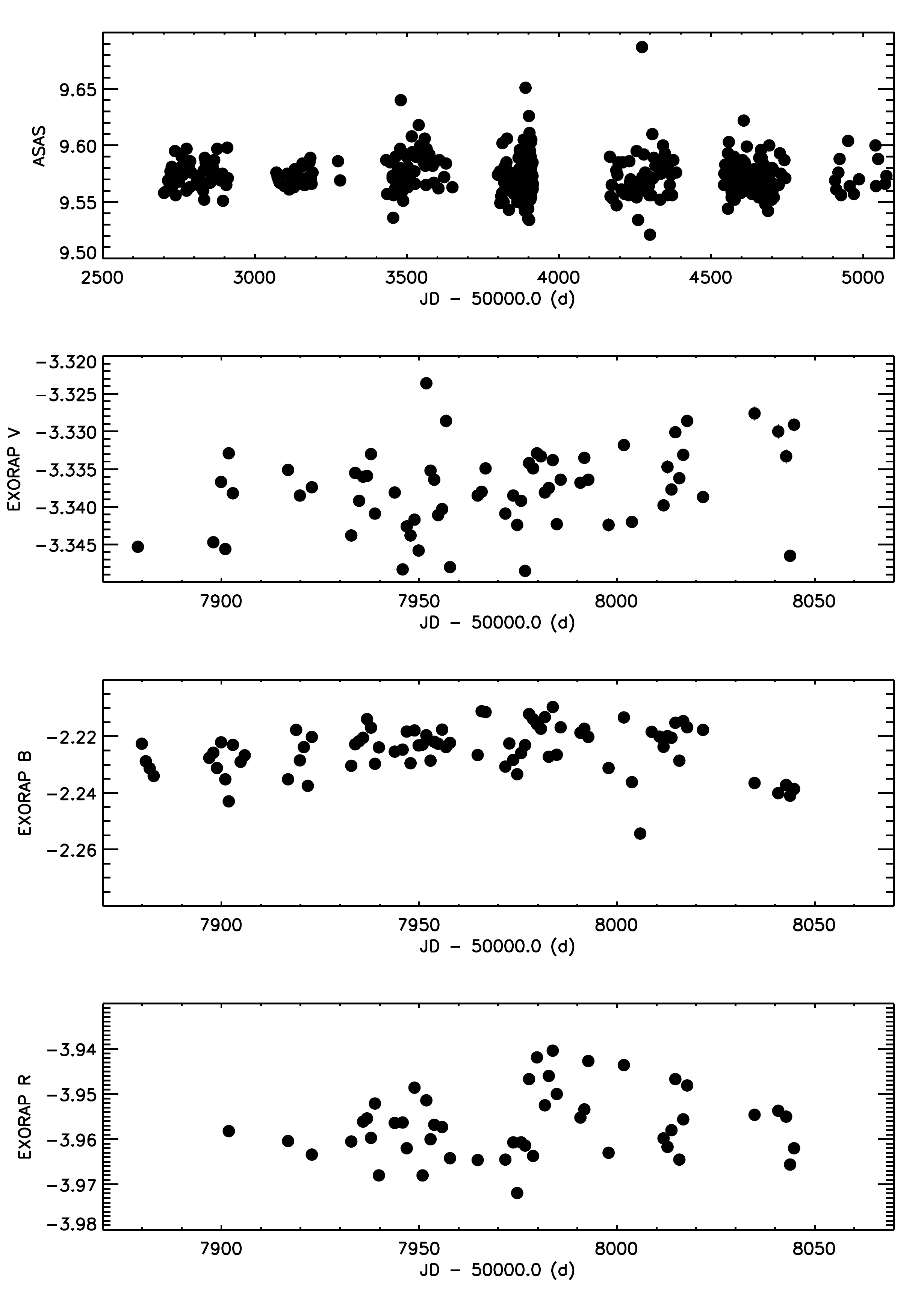}
\caption{Photometric time series: ASAS, EXORAP V, EXORAP B, EXORAP R, from top to bottom. The y axis indicates flux for ASAS data, differential magnitudes for EXORAP data.}
\label{fig:A2}
\end{figure*}

\begin{figure*}
\centering
\includegraphics[width=14cm]{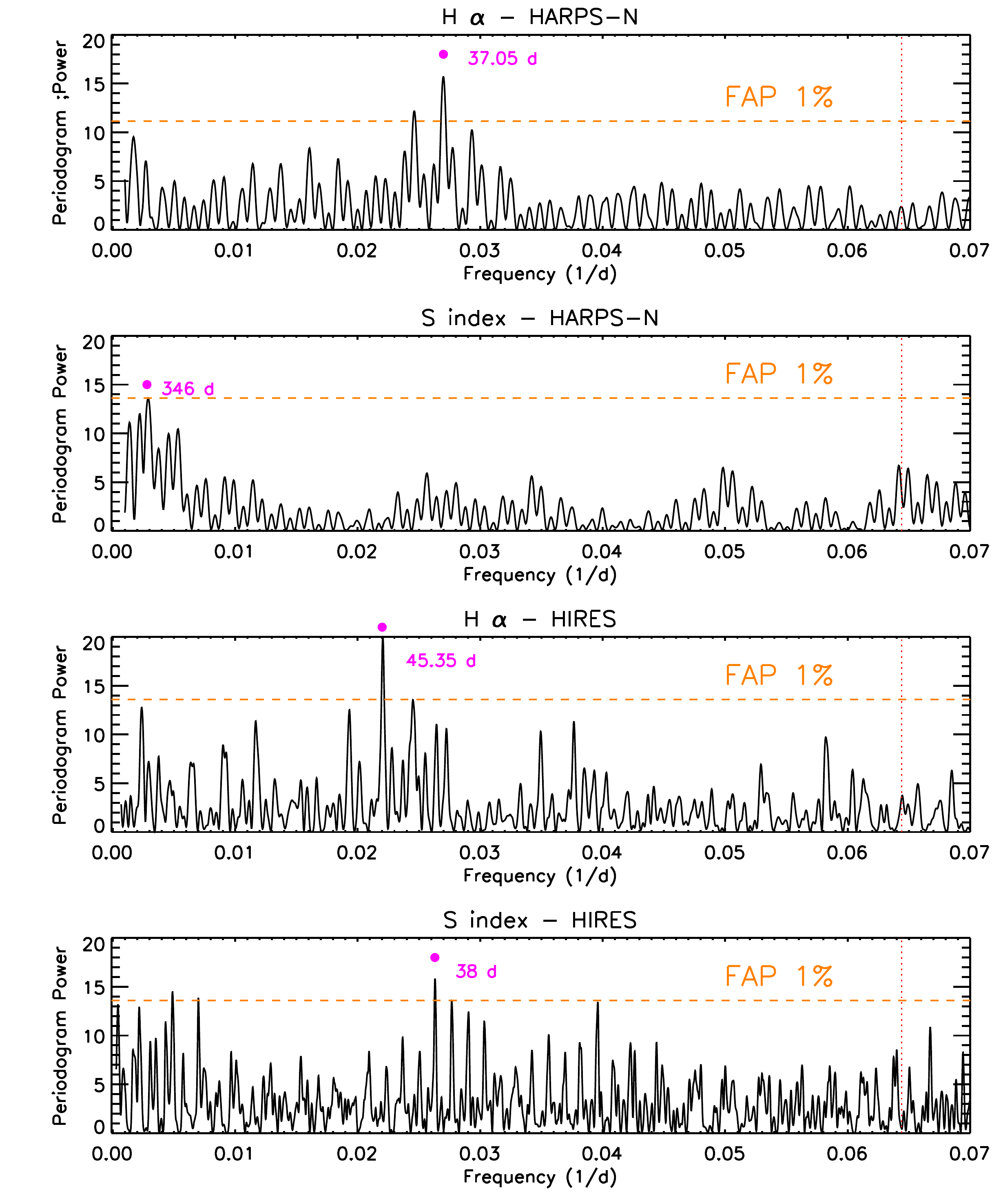}
\caption{GLS periodograms of activity indexes (from top to bottom): H$\alpha$ and Sindex (HARPS-N); H$\alpha$ and Sindex (HIRES). The 15.5 d period is indicated in each periodogram, as the red dashed line. The significant periods discussed in Sect.~\ref{3.4}, are indicated with a magenta dot. }
\label{fig:A3}
\end{figure*}

\begin{figure*}
\centering
\includegraphics[width=14cm]{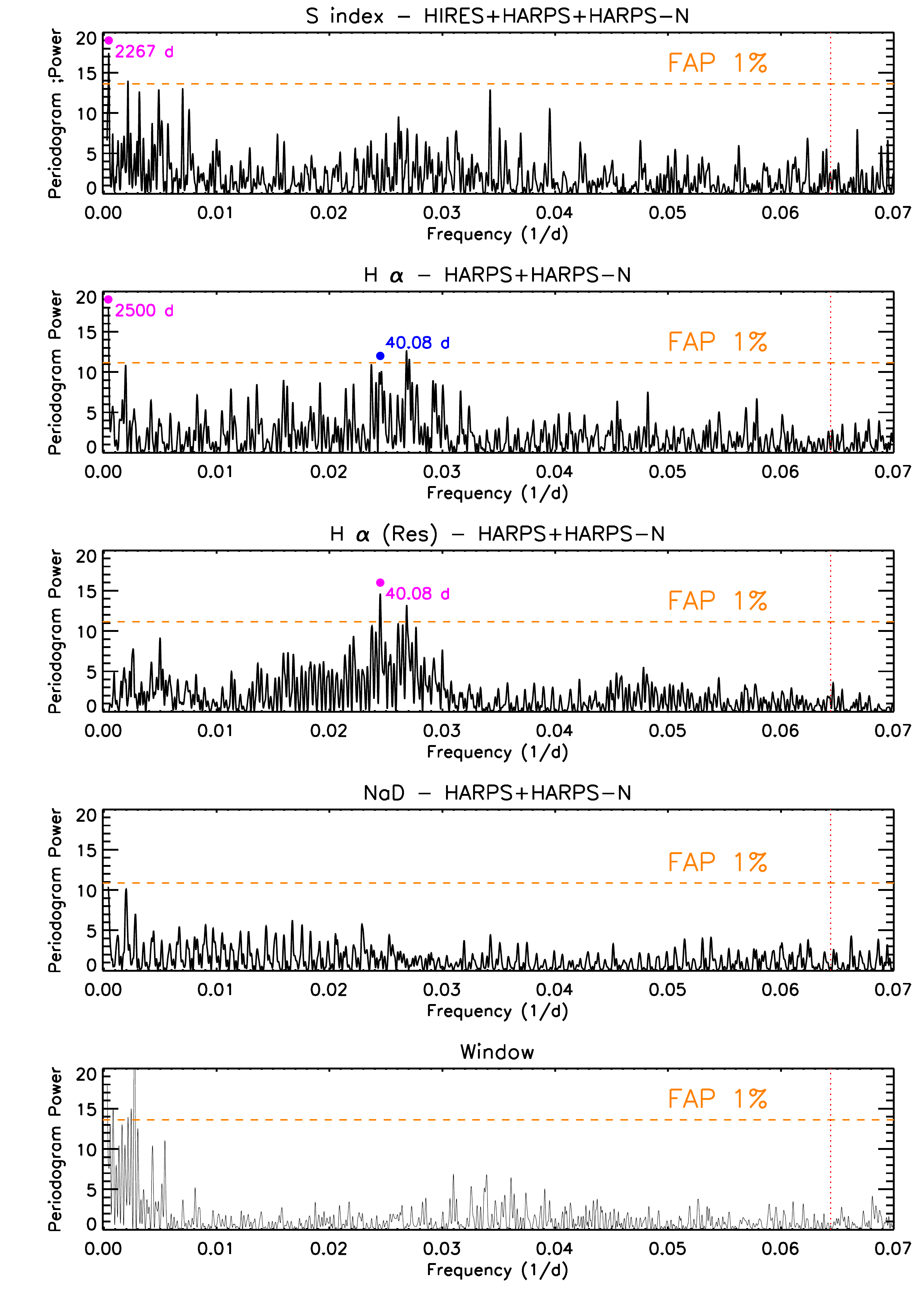}
\caption{GLS periodograms of activity indexes (from top to bottom): Sindex (HIRES+HARPS+HARPS-N); H$\alpha$ and NaD (HARPS+HARPS-N) and window function. The 15.5 d period is indicated in each periodogram, as the red dashed line. The significant periods discussed in Sect.~\ref{3.4}, are indicated with a magenta dot. In the H$\alpha$ periodogram (HARPS+HARPS-N), we also indicated, with a blue dot, the 40.8 d period which becomes significant in the residual time series, after the removal of the long 2500 d period. }
\label{fig:A4}
\end{figure*}

\begin{figure*}
\centering
\includegraphics[width=14cm]{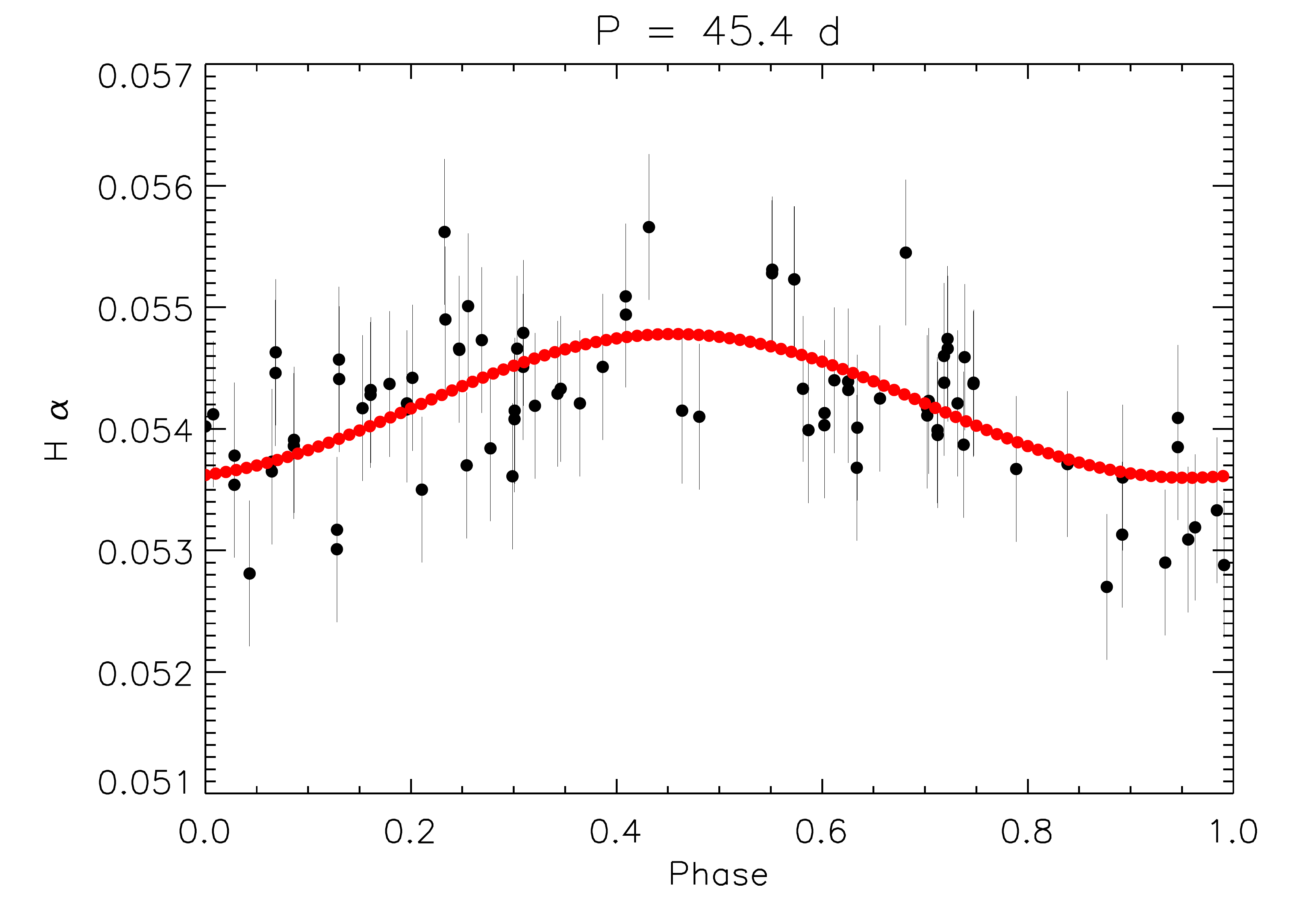}
\includegraphics[width=14cm]{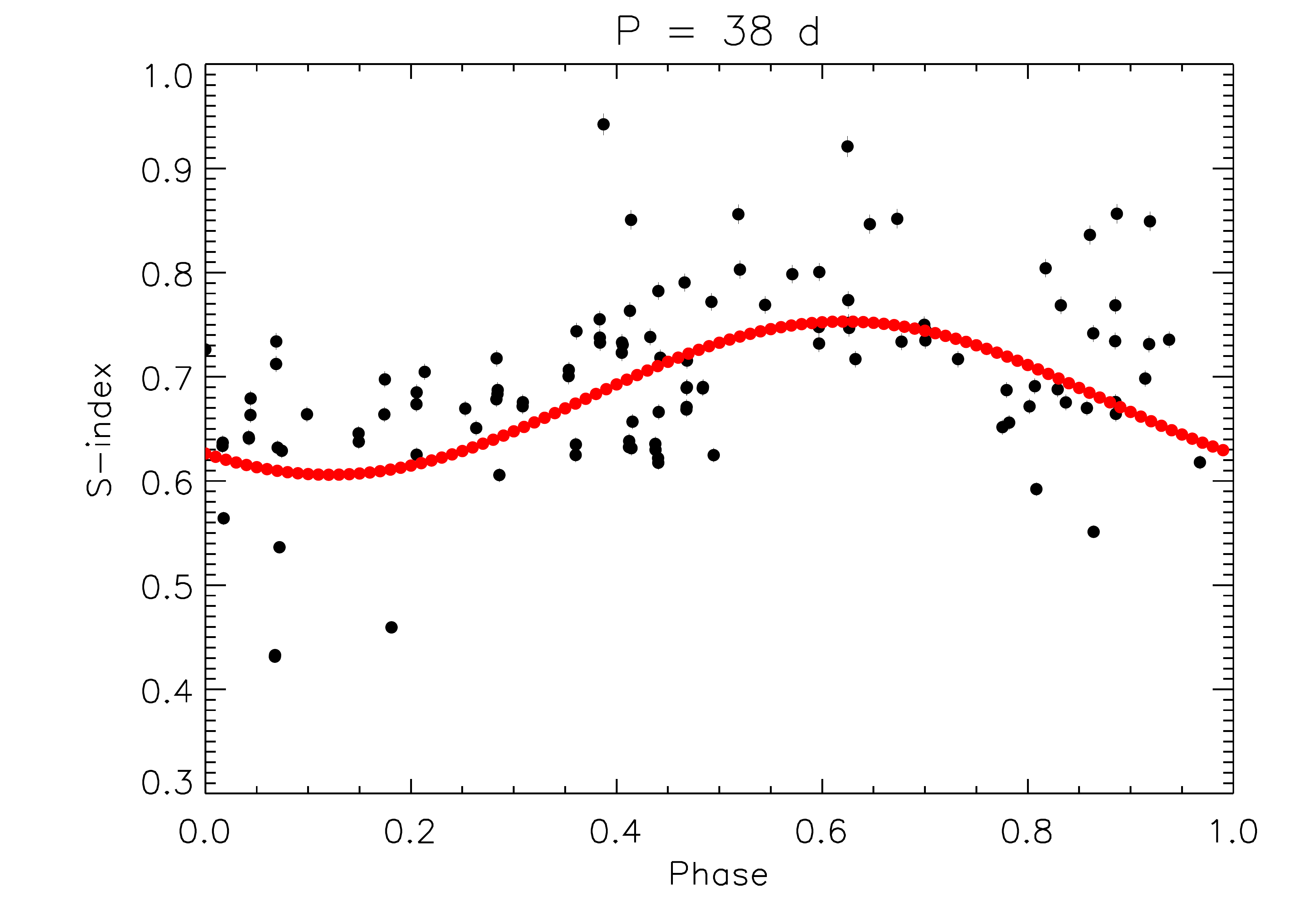}
\caption{Upper panel: H$\alpha$ data folded at the best period (HIRES, 114 data points). Lower panel: $S$-index data folded at the best period (HIRES, 114 data points). The red line is the best-fit solution in both panels.}
\label{fig:A5}
\end{figure*}

\begin{figure*}
\centering
\includegraphics[width=14cm]{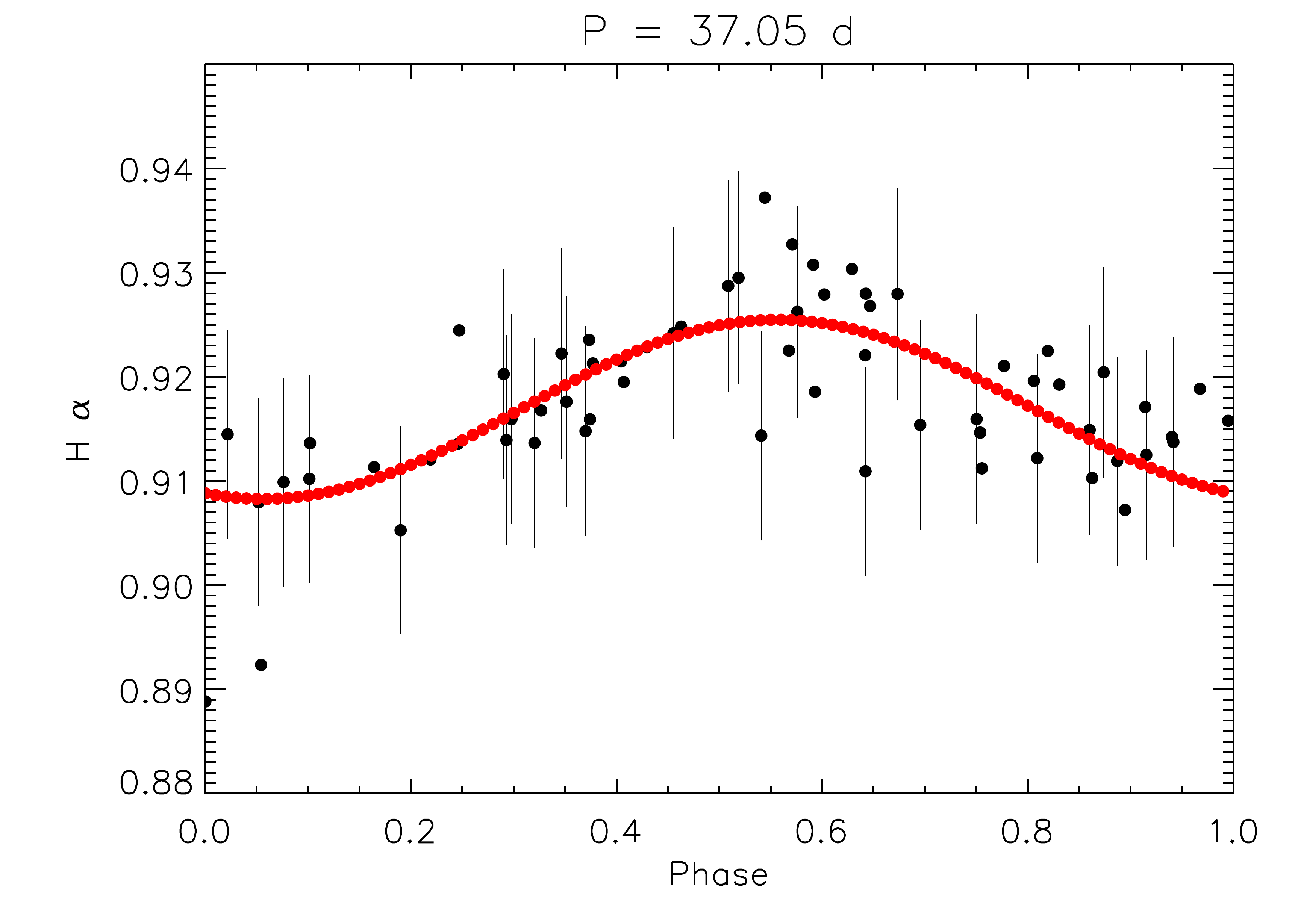}
\includegraphics[width=14cm]{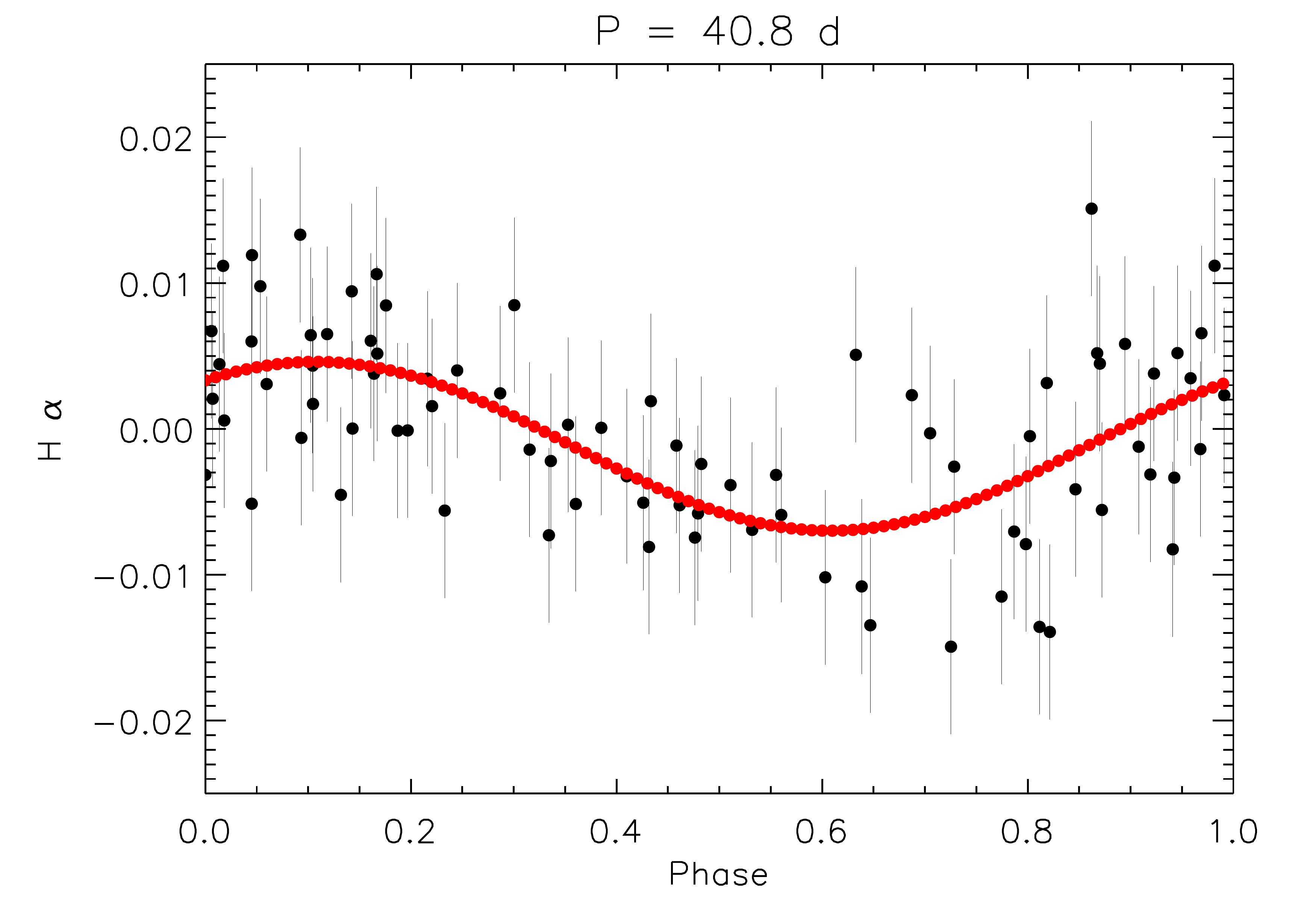}
\caption{Upper panel: H$\alpha$ data folded at the best period (HARPS-N, 64 data points). Lower panel: H$\alpha$ residual data (HARPS + HARPS-N, 84 data points), after the subtraction of the 2500 d period, folded at the best period. The red line is the best fit solution in both panels.}
\label{fig:A6}
\end{figure*}

\end{document}